%% file: quanto-cade.tex
\documentclass{llncs}

\usepackage{hyperref}
\usepackage{framed}

\usepackage{amsmath,amssymb}
\usepackage{xspace,enumerate,color,epsfig}
\usepackage{graphicx}
\graphicspath{{.}{./figures/}}
\usepackage{marvosym}

\usepackage{tikzfig}
\usepackage{keycommand}
\usepackage{bm}

\usepackage{wrapfig}

\newcommand{\smallify}{\tikzstyle{every picture}=[scale=0.14,every node/.style={scale=0.8},font=\footnotesize]
\tikzstyle{directed}=[decoration={markings,mark=at position 1 with {\arrow[scale=0.7]{>}}},postaction={decorate}]}

\newcommand{\unsmallify}{\tikzstyle{every picture}=[baseline=-0.25em,scale=0.35,every node/.style={},font=\footnotesize]
\tikzstyle{directed}=[decoration={markings,mark=at position 1 with {\arrow[scale=0.9]{>}}},postaction={decorate}]}

\usepackage[color]{changebar}

\tikzstyle{gn}=[circle,fill=green,draw=black, inner sep=0 pt, minimum size=0.16cm]
\tikzstyle{rn}=[circle,fill=red,draw=black, inner sep=0 pt, minimum size=0.16cm]
\tikzstyle{small black vdot}=[fill=black, minimum width=0pt,minimum height=0pt,draw,shape=circle, inner sep=0 pt, minimum size=0.08cm]
\tikzstyle{bbox}=[rectangle,fill=blue,draw=blue,scale=0.6]

\tikzstyle{wire}=[none]
\tikzstyle{X}=[rn]
\tikzstyle{Z}=[gn]
\tikzstyle{directed}=[decoration={markings,mark=at position 1 with {\arrow[scale=0.9]{>}}},postaction={decorate}]
\tikzstyle{simple}=[-,draw=black]

\tikzstyle{every picture}=[baseline=-0.25em,scale=0.35,font=\footnotesize]
\tikzstyle{dotpic}=[]
\tikzstyle{diredges}=[every to/.style={diredge}]
\tikzstyle{math matrix}=[matrix of math nodes,left delimiter=(,right delimiter=),inner sep=2pt,column sep=1em,row sep=0.5em,nodes={inner sep=0pt},text height=1.5ex, text depth=0.25ex]

\tikzstyle{inline text}=[text height=1.5ex, text depth=0.25ex,yshift=0.5mm]
\tikzstyle{label}=[font=\footnotesize,text height=1.5ex, text depth=0.25ex,yshift=0.5mm]
\tikzstyle{left label}=[label,anchor=east,xshift=1.5mm]
\tikzstyle{right label}=[label,anchor=west,xshift=-1.5mm]

\tikzstyle{empty diagram}=[draw=gray!40!white,dashed,shape=rectangle,minimum width=1cm,minimum height=1cm]
\tikzstyle{empty diagram small}=[draw=gray!50!white,dashed,shape=rectangle,minimum width=0.6cm,minimum height=0.5cm]

\tikzstyle{bbox edge}=[draw=blue]
\tikzstyle{fixed bbox edge}=[draw=red]
\tikzstyle{bbox include}=[->,draw=blue]
\tikzstyle{bbox corner}=[inner sep=0pt,rectangle,fill=blue,draw=blue,minimum width=1.5mm,minimum height=1.5mm]
\tikzstyle{fixed bbox corner}=[inner sep=0pt,rectangle,fill=red,draw=red,minimum width=1.5mm,minimum height=1.5mm]

\tikzstyle{dot}=[inner sep=0mm,minimum width=2mm,minimum height=2mm,draw,shape=circle,text depth=-0.1mm]

\tikzstyle{ddot}=[inner sep=0mm, doubled, minimum width=2.5mm,minimum height=2.5mm,draw,shape=circle]

\tikzstyle{black dot}=[dot,fill=black]
\tikzstyle{white dot}=[gn] 
\tikzstyle{green dot}=[white dot] 
\tikzstyle{gray dot}=[rn] 
\tikzstyle{red dot}=[gray dot] 

\tikzstyle{black ddot}=[ddot,fill=black]
\tikzstyle{white ddot}=[ddot,fill=white]
\tikzstyle{gray ddot}=[ddot,fill=gray!40!white]

\tikzstyle{gray edge}=[gray!40!white]

\tikzstyle{small dot}=[inner sep=0.5mm,minimum width=0pt,minimum height=0pt,draw,shape=circle]

\tikzstyle{small black dot}=[small dot,fill=black]
\tikzstyle{small white dot}=[small dot,fill=white]
\tikzstyle{small gray dot}=[small dot,fill=gray!40!white]

\tikzstyle{white rect ddot}=[draw=black,fill=white,doubled,minimum size=5mm,font=\footnotesize,rectangle,rounded corners=2.5mm,inner sep=0.2mm]
\tikzstyle{gray rect ddot}=[draw=black,fill=gray!40!white,doubled,minimum size=6mm,font=\footnotesize,rectangle,rounded corners=3mm]

\tikzstyle{small box}=[rectangle,inline text,fill=white,draw,minimum height=5mm,yshift=-0.5mm,minimum width=5mm,font=\small]
\tikzstyle{small gray box}=[small box,fill=gray!30]
\tikzstyle{medium box}=[rectangle,inline text,fill=white,draw,minimum height=5mm,yshift=-0.5mm,minimum width=10mm,font=\small]
\tikzstyle{square box}=[small box]
\tikzstyle{medium gray box}=[small box,fill=gray!30]
\tikzstyle{semilarge box}=[rectangle,inline text,fill=white,draw,minimum height=5mm,yshift=-0.5mm,minimum width=12.5mm,font=\small]
\tikzstyle{large box}=[rectangle,inline text,fill=white,draw,minimum height=5mm,yshift=-0.5mm,minimum width=15mm,font=\small]
\tikzstyle{large gray box}=[small box,fill=gray!30]

\tikzstyle{gray square point}=[small box,fill=gray!50]

\tikzstyle{dphase box white}=[dbox]
\tikzstyle{dphase box gray}=[dbox,fill=gray!50!white]

\tikzstyle{point}=[regular polygon,regular polygon sides=3,draw,scale=0.75,inner sep=-0.5pt,minimum width=9mm,fill=white,regular polygon rotate=180]
\tikzstyle{copoint}=[regular polygon,regular polygon sides=3,draw,scale=0.75,inner sep=-0.5pt,minimum width=9mm,fill=white]
\tikzstyle{dpoint}=[point,doubled]
\tikzstyle{dcopoint}=[copoint,doubled]

\tikzstyle{wide copoint}=[fill=white,draw,shape=isosceles triangle,shape border rotate=90,isosceles triangle stretches=true,inner sep=0pt,minimum width=1.5cm,minimum height=6.12mm]
\tikzstyle{wide point}=[fill=white,draw,shape=isosceles triangle,shape border rotate=-90,isosceles triangle stretches=true,inner sep=0pt,minimum width=1.5cm,minimum height=6.12mm,yshift=-0.0mm]
\tikzstyle{wide point plus}=[fill=white,draw,shape=isosceles triangle,shape border rotate=-90,isosceles triangle stretches=true,inner sep=0pt,minimum width=1.74cm,minimum height=7mm,yshift=-0.0mm]

\tikzstyle{wide dpoint}=[fill=white,doubled,draw,shape=isosceles triangle,shape border rotate=-90,isosceles triangle stretches=true,inner sep=0pt,minimum width=1.5cm,minimum height=6.12mm,yshift=-0.0mm]

\tikzstyle{tinypoint}=[regular polygon,regular polygon sides=3,draw,scale=0.55,inner sep=-0.15pt,minimum width=6mm,fill=white,regular polygon rotate=180] 

\tikzstyle{white point}=[point]
\tikzstyle{white dpoint}=[dpoint]
\tikzstyle{green point}=[white point] 
\tikzstyle{white copoint}=[copoint]
\tikzstyle{gray point}=[point,fill=gray!40!white]
\tikzstyle{gray dpoint}=[gray point,doubled]
\tikzstyle{red point}=[gray point] 
\tikzstyle{gray copoint}=[copoint,fill=gray!40!white]
\tikzstyle{gray dcopoint}=[gray copoint,doubled]

\tikzstyle{black point}=[point,fill=black]
\tikzstyle{black copoint}=[copoint,fill=black]

\tikzstyle{tiny gray point}=[tinypoint,fill=gray!40!white]

\tikzstyle{diredge}=[->]
\tikzstyle{rdiredge}=[<-]
\tikzstyle{thickdiredge}=[->, very thick]
\tikzstyle{pointer edge}=[->,very thick,gray]
\tikzstyle{pointer edge part}=[very thick,gray]
\tikzstyle{dashed edge}=[gray!70!white]

\pgfkeyssetvalue{/tikz/shorten left}{0pt}
\pgfkeyssetvalue{/tikz/shorten right}{0pt}

\title{Quantomatic: A Proof Assistant for \\ Diagrammatic Reasoning
\thanks{The final publication is available at Springer via
\href{http://dx.doi.org/10.1007/978-3-319-21401-6_22}
{http://dx.doi.org/10.1007/978-3-319-21401-6\_22}}
}

\titlerunning{Quantomatic}

\author{Aleks Kissinger \and Vladimir Zamdzhiev}
\authorrunning{A. Kissinger & V. Zamdzhiev}
\institute{University of Oxford\\
\email{\{aleks.kissinger|vladimir.zamdzhiev\}@cs.ox.ac.uk}
}

\begin{document}

\maketitle

\begin{abstract}
  Monoidal algebraic structures consist of operations that can have multiple outputs as well as multiple inputs, which have applications in many areas including categorical algebra, programming language semantics, representation theory, algebraic quantum information, and quantum groups. String diagrams provide a convenient graphical syntax for reasoning formally about such structures, while avoiding many of the technical challenges of a term-based approach. Quantomatic is a tool that supports the (semi-)automatic construction of equational proofs using string diagrams. We briefly outline the theoretical basis of Quantomatic's rewriting engine, then give an overview of the core features and architecture and give a simple example project that computes normal forms for commutative bialgebras.
\end{abstract}

\section{Introduction}

Quantomatic is a graphical proof assistant. Rather than using terms as the primitive objects in proofs, it uses \textit{string diagrams}. String diagrams provide a simple way of expressing collections of maps or processes that have been plugged together. They consist of boxes representing the processes, and (typed) wires connecting them. Wires are allowed to be open (i.e. not connected to a box) at one or both ends, giving a notion of \textit{input} and \textit{output} for a string diagram.

\begin{wrapfigure}{l}{0.3\textwidth}
  \vspace{-10pt}
  \centering
  \input{compound-process-new.tikz}
  \caption{A string diagram}
  \vspace{-20pt}
\end{wrapfigure}
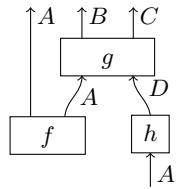
String diagram notation was first used by Penrose~\cite{Penrose} as an alternative to tensor notation for applications in theoretical physics. In 1991, Joyal and Street showed that string diagrams were actually much more general~\cite{JS}, serving to not just represent tensors, but morphisms in any monoidal category. In other words, it is possible to reason about any collection of processes or maps that has well-behaved parallel and sequential composition operations (usually written $\otimes$ and $\circ$, respectively) using string diagrams. This includes familiar examples such as functions (where $\otimes := \times$ is just the Cartesian product), and other non-Cartesian examples such as multi-linear maps or matrices over a semi-ring (where $\otimes$ is a genuine tensor product).

Recently, there has been much interest in diagrammatic theories in a wide variety of areas such as Petri nets~\cite{Sobocinski:2010aa}, programming language semantics~\cite{Mellies:2014aa}, natural language processing~\cite{Bob-Coecke:2010fk}, systems biology~\cite{DanosEtAl-LICS2010}, control theory~\cite{Baez2014a,Bonchi2015}, program parallelisation~\cite{gudmund:11c}, and in interactive theorem proving~\cite{tinker14}. It has also played a major role in categorical quantum mechanics~\cite{AC2004}. In particular, the string diagram-based \textit{ZX-calculus}~\cite{CD2} has found numerous applications within quantum computing (see e.g.~\cite{Duncan:2013lr,Hillebrand:2011qy}). String diagrammatic reasoning has also produced results which had been previously unknown~\cite{mbqc_circuit,mehrnoosh_experiment}.

The current version of Quantomatic supports the construction of derivations, which are transitive chains of diagram rewrites, as well as simple mechanisms for automated simplification of diagrams and lemma/theorem export and re-use. The theoretical foundations of Quantomatic have been described in previous papers~\cite{DK,pattern_graphs,Merry:2013aa} and this paper is the first system description of the Quantomatic software itself. After introducing the main concepts of diagrammatic reasoning in Section~\ref{sec:diagrams}, we describe how Quantomatic builds derivations and how those derivations can be included in papers or shared in the web in Section~\ref{sec:quanto_basics}. We show how to implement simplification strategies using a simple combinator language in Section~\ref{sec:simp_strategies} and describe an example project involving bialgebras in Section~\ref{sec:bialgebra}. Then, we give an overview of the architecture of the system in Section~\ref{sec:architecture}, and show how it can be extended with new graphical theories. We give details on obtaining Quantomatic and discuss related and future work in Section~\ref{sec:future_work}.

\section{Diagrammatic Reasoning}\label{sec:diagrams}

String diagram rewriting can be seen as a generalisation of (linear) term rewriting.\footnote{Non-linear term rewriting can be encoded by introducing special `copy' and `delete' nodes which obey certain naturality conditions. However, when $\otimes \neq \times$, these don't exist in general.} We can see how this works via a simple example. A commutative monoid is a set $A$, along with a binary operation $(- \cdot -)$ and a constant $e \in A$ such that:
\begin{equation}\label{eq:monoid}
  (a \cdot b) \cdot c = a \cdot (b \cdot c) \qquad\qquad
  a \cdot e = a = e \cdot a \qquad\qquad
  a \cdot b = b \cdot a
\end{equation}
Naturally, we can treat these equations as term rewrite rules, with free variables $a, b, c$. To apply a rule, we instantiate the free variables, then use it to replace a sub-term. For example, the assignment $\{ a := x, b := y \cdot e, c := z \}$ in the first rule yields $(x \cdot (y \cdot e)) \cdot z = x \cdot ( (y \cdot e) \cdot z)$, which could be applied in, e.g.:
\begin{equation}\label{eq:subst}
  w \cdot \bm{((x \cdot (y \cdot e)) \cdot z)} =
  w \cdot \bm{(x \cdot ( (y \cdot e) \cdot z))}
\end{equation}
We could express the same thing by rewriting string diagrams, which in this case are just trees. Representing $\cdot$ as a node with two inputs and one output and $e$ as a node with just one output, the equations~\eqref{eq:monoid} become:
\begin{equation}\label{eq:diag-monoid}
  \scalebox{0.9}{\input{assoc_w_points.tikz} \qquad \qquad
  \input{unit_w_points.tikz} \qquad\qquad
  \input{comm_w_points.tikz}}
\end{equation}
In fact, the variable names on the inputs are no longer necessary. The role of the variables is played by the fact that the LHS and the RHS share a common boundary. That is, there is a 1-to-1 correspondence between inputs/outputs on the LHS and those on the RHS.
The substitution~\eqref{eq:subst} can then be depicted simply as cutting out the LHS of this rule and gluing in the RHS, using the shared boundary:
\begin{equation}\label{eq:assoc-rewrite}
  \scalebox{0.9}{\input{assoc_rewrite.tikz}}
\end{equation}
The benefit of this approach is that it treats inputs and outputs symmetrically. For instance, we can define a \textit{cocommutative comonoid} by simply flipping the generators and equations upside-down:
\[ \scalebox{0.9}{\input{coassoc.tikz} \qquad\qquad
   \input{counit-law.tikz} \qquad\qquad
   \input{cocomm.tikz}} \]
Many interesting and useful structures arise by letting algebraic structures like monoids interact with their `coalgebraic' counterparts. For example, a \textit{commutative bialgebra} consists of a commutative monoid, a cocommutative comonoid, and three rules governing their interaction:
\begin{equation}\label{eq:bialg}
  \scalebox{0.9}{\input{bialgebra.tikz} \qquad\qquad
  \input{white-copy.tikz} \qquad\qquad
  \input{gray-cocopy.tikz}}
\end{equation}
Rewriting with general diagrams proceeds just like the tree rewriting above:
\begin{center}
  \scalebox{0.9}{\input{bialg_app.tikz}}
\end{center}
This process of cutting out the LHS and gluing in the RHS along a shared boundary is called \textit{double-pushout (DPO) graph rewriting}. The precise formulation of DPO rewriting for string diagrams is provided in~\cite{DK}.

From hence forth, we will assume all nodes are commutative and cocommutative, or in other words, nodes are invariant to permutations of their inputs and outputs. A current limitation of Quantomatic is that it does not maintain an ordering on inputs/outputs for individual nodes, so this is true by default. A semantics for diagrams with non-commutative nodes is detailed in~\cite{Noncomm-bb}, but is not yet implemented (see Section~\ref{sec:future_work}).

One of the unique aspects of Quantomatic is that it supports a graphical pattern syntax called \textit{!-box notation} for expressing infinite families of rules, typically involving variable-arity generators. For example, we could alternatively define commutative monoids using $n$-ary multiplication operations, subject to the rules that adjacent multiplications merge and the `1-ary multiplication' does nothing:
\begin{equation}\label{eq:nary-merge}
  \scalebox{0.9}{\input{nary-merge.tikz} \qquad\qquad \input{id-to-wire.tikz}}
\end{equation}
One could recursively define these $n$-ary multiplications as (e.g. left-associated) trees of binary multiplications, where a `$0$-ary multiplication' is just the unit. Then, by associativity, commutativity, and unit laws, any two trees with the same number of inputs will be equal, from which the two equations above follow.

To represent repetition, we can enclose certain parts of the diagram in !-boxes, which indicate that the marked sub-diagram (along with any wires in or out) can be duplicated any number of times.
Replacing the ellipses with !-boxes in~\eqref{eq:nary-merge} yields:
\begin{equation}\label{eq:tree-merge-bb}
  \scalebox{0.9}{\input{bb-tree-merge.tikz} \qquad\qquad \input{id-to-wire.tikz}}
\end{equation}
An instance of this rule effectively amounts to fixing the number times the contents of $b$ and $c$ are repeated. In order to ensure that all instances are valid string diagram rules (i.e. they share a common boundary), !-box rules must satisfy two well-formedness conditions: (i) the !-boxes on both sides are in bijective correspondance indicated by their labels, and (ii) an input (resp. output) is in a !-box $b$ on the LHS if and only if it is also in $b$ on the RHS, where pairs of inputs or outputs are again indicated by their labels. !-boxes can also be nested in each other, which adds one additional condition, but for simplicity we will ignore this case. More details on !-boxesand their formal semantics can be found in~\cite{pattern_graphs}.

\begin{figure}[b]
\centering
\vspace{-15pt}
\includegraphics[width=0.8\textwidth]{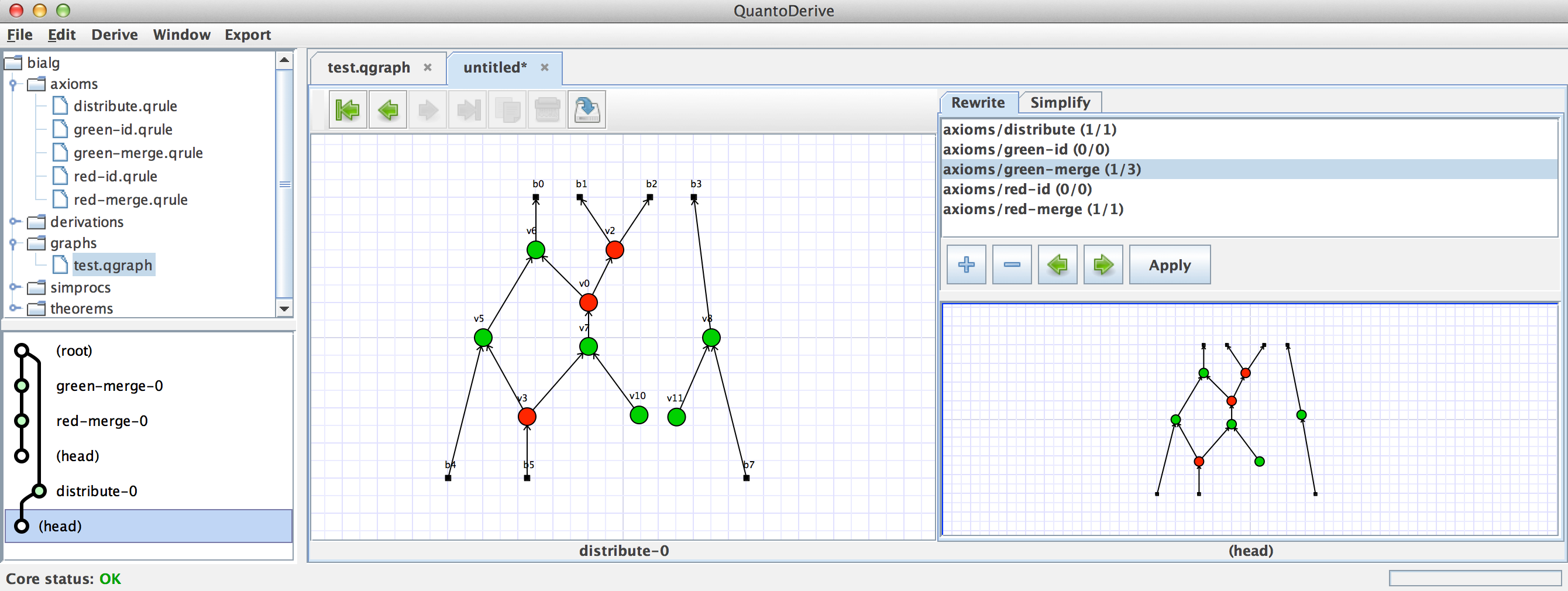}
\caption{\label{fig:derivation} Derivation editor in Quantomatic}
\vspace{-15pt}
\end{figure}

\section{Constructing Proofs in Quantomatic}\label{sec:quanto_basics}

Quantomatic allows a user to define a set of diagram equations and use them to prove theorems by means of \textit{derivations}. A derivation is simply a transitive chain of rewrite steps, using axioms or other theorems within the project. To begin working in Quantomatic, the user creates a project based on a \textit{graphical theory}, which defines the kinds of admissible nodes in a diagram and how they should be presented to the user (see Section~\ref{sec:architecture}). At this point, they can define some axioms, i.e.~diagram equations (possibly containing !-boxes) subject to the well-formedness conditions listed at the end of Section~\ref{sec:diagrams}.

After a set of axioms is defined, they can be used in a derivation. First, the user creates a new graph using the graph editor and chooses to start a new derivation from the menu. The user is then presented with the derivation editor, which is used to explore the derivation history or extend it by applying rewrite rules. The history view on the left shows a chain of proof steps. The history can also be branched off at any step, allowing the user to explore multiple (possibly failed) rewriting paths on the way to producing a proof.

The nodes in this tree are organised into two categories: \textit{proof steps} and \textit{proof heads}. The former represent the application of a rewrite rule. With a proof step selected, the user sees the before and after graphs side-by-side, with the changed portion highlighted. The user can grow the derivation from a proof head. Here, they see the current graph next to a series of controls (as in Fig.~\ref{fig:derivation}). If the `Rewrite' panel is active, Quantomatic will eagerly look for matches of any active rewrite rules on the selected part of the graph on the left. This search is done in parallel, which is especially effective on multi-core machines at providing the desired rule application as soon as possible. Applying a rule will generate a new proof step and advance the proof head. The `Simplify' panel gives the user access to simplification procedures (see Section~\ref{sec:simp_strategies}), which will automatically produce proof steps until either the procedure terminates or is interrupted by the user. Once a derivation is complete, it can be exported as a new theorem, which is linked to the derivation and can be used in other derivations.

\begin{figure}[t]
  \centering

 \smallify
\fbox{\scalebox{0.5}{\parbox{0.57\textwidth}{
  \vspace{-3mm}
  \begin{theorem}
    \input{quanto-output-thm.tikz}
  \end{theorem}
  \vspace{-5mm}
  \begin{proof}\ \\
    \input{quanto-output.tikz}
  \end{proof}
  \vspace{-3mm}
}}}
\unsmallify \ \ \ 
\raisebox{-1.6cm}{\includegraphics[height=3.5cm]{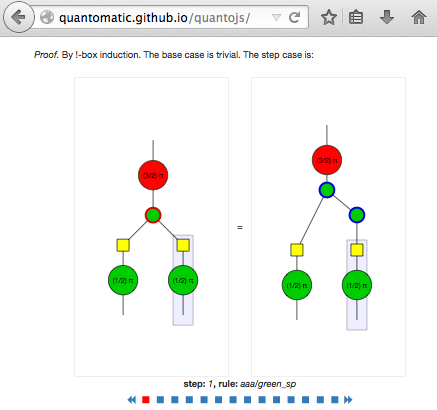}}

  \caption{\label{fig:export} \LaTeX{} and interactive HTML5 output from Quantomatic}
\end{figure}

One of the major advantages of diagrammatic reasoning is that it can produce nice, human-readable proofs. Proofs produced by Quantomatic can be shared in two ways. Graphs, rules, and derivations can be exported as \LaTeX{} and \verb!\input! directly in to papers (Fig.~\ref{fig:export}, left). The graphs are rendered using the PGF/Ti\textit{k}Z package, and are compatible with graphical editor Ti\textit{k}ZiT, in case further manual tweaking is required. It is also possible to embed graphs, rules, and derivations from a Quantomatic project in HTML5 using \texttt{Quanto.js}. After linking to a Quantomatic project with a \verb!<meta>! tag, this script will substitute specially marked-up \verb!<div>! tags for interactive graphical views of proofs, rendered using \texttt{d3.js} (Fig.~\ref{fig:export}, right).

\section{Simplification Procedures}\label{sec:simp_strategies}

Quantomatic allows for custom simplification procedures (simprocs). These are functions implemented in Poly/ML which send a graph to a lazy sequence of proof steps, which contain the name of the axiom/theorem used, the instantiated rewrite rule, and the rewritten graph. Simprocs are then registered with the Quantomatic GUI by calling \texttt{register\_simproc}. When a simproc is invoked in the derivation editor, it is passed the current graph, and proof steps are pulled one at a time until either the sequence is exhausted or the user cancels simplification. To construct simprocs, Quantomatic provides a combinator language:
\begin{verbatim}
                ++ :: simproc * simproc -> simproc
              LOOP :: simproc -> simproc
            REDUCE :: rule -> simproc
        REDUCE_ALL :: ruleset -> simproc
      REDUCE_WHILE :: (graph -> bool) -> rule -> simproc
       type metric := graph -> int
  REDUCE_METRIC_TO :: int -> metric -> simproc
\end{verbatim}

\begin{wrapfigure}{r}{0.45\textwidth}
  \vspace{-20pt}
  \includegraphics[width=0.45\textwidth]{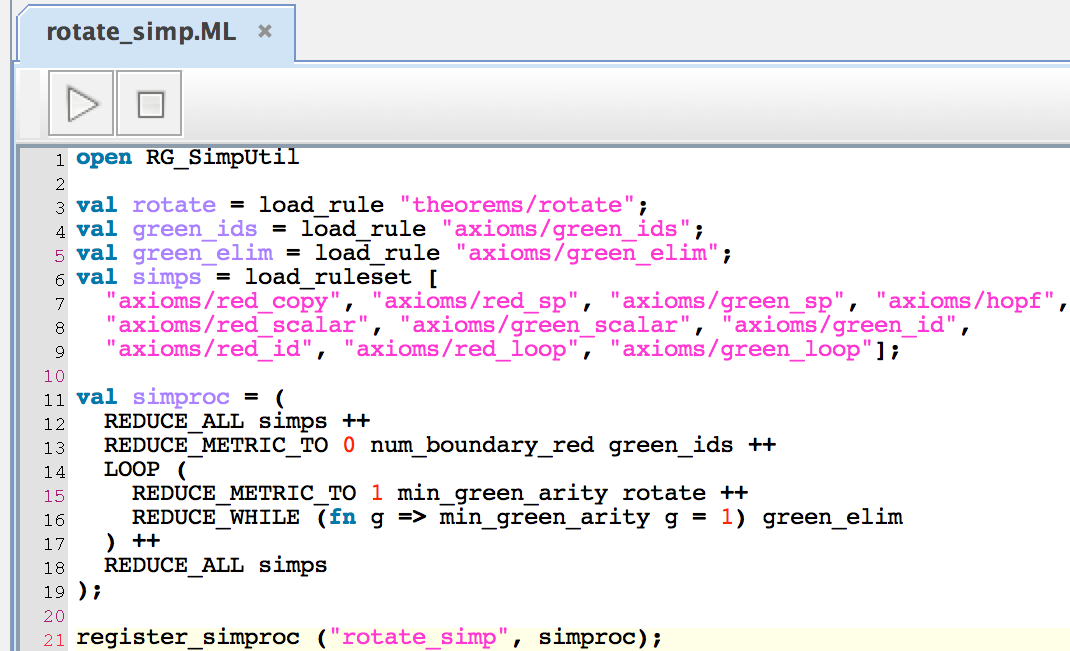}
  \caption{\label{fig:simproc} A simproc in Quantomatic}
  \vspace{-20pt}
\end{wrapfigure}
The combinator \texttt{++} will chain the last graph produced by the first simproc into the second simproc. \texttt{LOOP} will repeatedly chain a simproc into itself, until the simproc produces no new proof steps. \texttt{REDUCE} will repeatedly apply the first matching of the given rule, and \texttt{REDUCE\_ALL} does the same, but takes a set of rules. \texttt{REDUCE\_WHILE} will keep reducing as long as the graph satisfies the given pre-condition. \texttt{REDUCE\_METRIC\_TO} is useful for using non-terminating rules in strategies. It takes an integer $k$ and a function $m$. It will then repeatedly apply the given rule to a graph $g$ as long as $m(g) > k$ and $m(g)$ is reduced by the rule application.

For terminating, confluent rewrite systems, a single call to \texttt{REDUCE\_ALL} will usually suffice. However, strategies are very useful for more ill-behaved systems. For example, Figure~\ref{fig:simproc} shows a simproc that computes pseudo-normal forms for the theory of interacting bialgebras described in~\cite{pawel-bialg}, which currently has no known convergent completion.

\section{Example Project: Bialgebras}\label{sec:bialgebra}

As mentioned in Section~\ref{sec:diagrams}, a bialgebra consists of a monoid and a comonoid, subject to three extra equations~\eqref{eq:bialg}. There is also a more efficient way to define commutative bialgebras, following the $n$-ary presentation of monoids described in Section~\ref{sec:diagrams}. A commutative bialgebra can be presented in terms of an $n$-ary multiplication and $n$-ary comultiplication, subject to the monoid `tree-merge' rules in~\eqref{eq:tree-merge-bb}, as well as the comonoid versions:
\begin{equation}\label{eq:cotree-merge}
  \scalebox{0.9}{\input{bb-tree-merge-gray.tikz} \qquad\qquad
\input{id-to-wire-gray.tikz}}
\end{equation}
and one additional rule. Whenever an $n$-ary multiplication meets an $m$-ary comultiplication, the two nodes can be replaced by a complete bipartite graph:
\begin{equation}\label{eq:gen-bialg}
  \scalebox{0.9}{$\input{gen-bialg.tikz} \qquad\leadsto\qquad \input{gen-bialg-bbox.tikz}$}
\end{equation}
The 5 equations depicted in \eqref{eq:tree-merge-bb}, \eqref{eq:cotree-merge}, and \eqref{eq:gen-bialg} can be added to a Quantomatic project. Since they are strongly normalising, the following na\"ive strategy will compute normal forms:
\begin{verbatim}
  val simps = load_ruleset ["axioms/red-merge", "axioms/red-id",
    "axioms/green-merge", "axioms/green-id", "axioms/distribute"];
  register_simproc ("basic_simp", REDUCE_ALL simps);
\end{verbatim}

This bialgebra example is a small fragment of the ZX-calculus, which has about 20 basic rules and necessitates non-na\"ive simplification strategies. The bialgebra example and the ZX-calculus are available as projects on {\color{blue} \href{http://quantomatic.github.io}{\texttt{quantomatic.github.io}}}.

\section{Architecture}\label{sec:architecture}

\begin{figure}
\vspace{-5mm}
\centering
\scalebox{0.8}{\input{arch.tikz}}
\caption{Architecture of Quantomatic}\label{fig:quanto_arch}
\vspace{-5mm}
\end{figure}
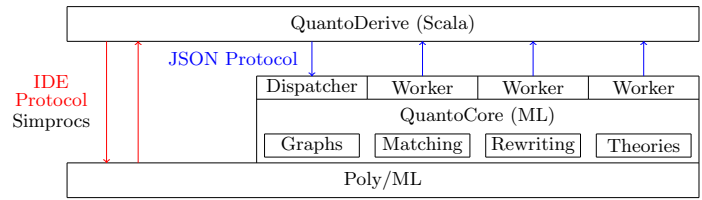

Quantomatic consists of two components: a reasoning engine written in Poly/ML called QuantoCore, and a GUI front-end written in Scala called QuantoDerive. QuantoCore handles matching and rewriting of diagrams, and can be extended via graphical theories. The GUI communicates to the core via a JSON protocol, which spawns independent workers to handle individual matching and rewriting requests. This allows the eager, parallel matching described in Section~\ref{sec:quanto_basics}. The GUI also communicates directly to Poly/ML using its built-in IDE protocol to register new simprocs written in ML. The core itself can be run in stand-alone mode or within Isabelle/ML. It forms the basis of two other graph-rewriting projects: QuantoCoSy~\cite{QuantoCosy}, which generates new graphical theories using conjecture synthesis (cf.~\cite{IsaCosy}), and Tinker~\cite{tinker14}, which implements a graphical proof strategy language for Isabelle and ProofPower.

Quantomatic is very flexible in terms of the data it can hold on nodes and edges. This can be something as simple as an enumerated type (e.g. a colour), standard types like strings and integers, or more complicated data like linear polynomials, lambda terms, or even full-blown programs. The specification of this data, along with how it should be unified during matching and displayed to the user, is called a \textit{graphical theory}. A graphical theory consists of two parts: a \texttt{.qtheory} file loaded into the GUI and an ML structure loaded into the core. The \texttt{.qtheory} is a JSON file used to register a new theory with the GUI, and provides basic information such as how node/edge data should be displayed to the user.

The ML structure provides four types, which Quantomatic treats as black boxes: \texttt{nvdata}, \texttt{edata}, \texttt{psubst}, and \texttt{subst}. The first two contain node data and edge data, respectively. The third type is for \textit{partial substitutions}, which are used to accumulate state during the course of matching one diagram against another. The fourth type is for \textit{substitutions}, which are partial substitutions that have been completed, or `solved', after matching is done. Quantomatic accesses these types using several hooks implemented in by theory:
\begin{verbatim}
     match_nvdata :: nvdata * nvdata -> psubst -> psubst option
      match_edata :: edata * edata -> psubst -> psubst option
     solve_psubst :: psubst -> [subst]
  subst_in_nvdata :: subst -> nvdata -> nvdata
   subst_in_edata :: subst -> edata -> edata
\end{verbatim}
\noindent The first two hooks are called every time a new node or edge is matched by the graph rewriting engine. The first argument is a pair consisting of the data on the pattern node (resp. edge) and the data on the target node (resp. edge). If the data matches successfully, any updates such as variable instantiations or new constraints are added to the \texttt{psubst}. If it fails (e.g. by introducing unsatisfiable constraints), the function returns \texttt{NONE}. Once matching is done, \texttt{solve\_psubst} is invoked to turn the accumulated constraints into an actual instantiation of node/edge data. Since we don't require node/edge data to have most-general unifiers, this is allowed to (lazily) return multiple solutions in general. The final two hooks are used to perform the instantiation of node/edge data on a rewrite rule. Once the theory provides these and a couple of other routine functions (e.g.~for (de)serialising data), QuantoCore handles the rest.

\section{Availability, Related, and Future Work}\label{sec:future_work}

Quantomatic is Free and Open Source Software, licensed under GPLv3. The project is hosted by Github, and source code and binaries for GNU/Linux, Mac OSX, and Windows are available from:
{\color{blue} \href{http://quantomatic.github.io}{\texttt{quantomatic.github.io}}}.
Example projects from Section~\ref{eq:bialg} are also available from the website. A page showing some of the features of \texttt{Quanto.js} is available at: {\color{blue} \href{http://quantomatic.github.io}{\texttt{quantomatic.github.io/quantojs}}}.

There are many tools for graph transformation, Quantomatic is unique in that it is designed specifically for diagrammatic reasoning. In other words, it is a general-purpose proof assistant for string-diagram based theories. Perhaps its closest relatives are general-purpose graph rewriting tools. GROOVE~\cite{groove} is a tool for graph transformation whose main focus is model checking of object-oriented systems. Like Quantomatic, GROOVE has a mechanism for specifying rules that can match many different concrete graphs, namely \textit{quantified graph transformation rules}. Other graph rewriting tools such as PROGRESS~\cite{progress} and AGG~\cite{agg} also have mechanisms that can be used to control application of a single rule to many concrete graphs. All of these mechanisms have quite different semantics from !-box rewriting, owing to the fact that the latter is specifically designed for transforming string diagrams. Its an open question whether any of these mechanisms could encode !-boxes.

There are three major directions in which we hope to extend Quantomatic. The first is in the support of non-commutative vertices and theories. The theoretical foundation for non-commutative graphical theories with !-boxes was given in~\cite{Noncomm-bb}. A big advantage of this is the ability to define new nodes which could be substituted for entire diagrams. This would allow extension of a theory by arbitrary, possibly recursive definitions. Secondly, we aim to go beyond `derivation-style' proofs into proper, goal-based backward reasoning. In~\cite{Kissinger:2012lr}, we introduced the concept of !-box induction, which was subsequently formalised~\cite{Merry:2013aa}. In conjunction with recursive definitions, this gives a powerful mechanism for introducing new !-box equations. This would also be beneficial even for purely equational proofs, as it is sometimes difficult to coax Quantomatic into performing the correct rewrite step in the presence of too much symmetry. It is also an important stepping stone toward providing QuantoCore with a genuine LCF-style proof kernel. Another, possibly complementary, approach is to integrate Quantomatic with an existing theorem prover, essentially as a `heavyweight tactic' for the underlying formal semantics of the diagram. In~\cite{Noncomm-bb}, this semantics is presented as a term language with wires as bound pairs of names, and we have had some preliminary success in formalising this language in Nominal Isabelle. Third, it was recently shown in \cite{trivial_overlap} that placing a natural restriction on !-boxes yields a proper subset of context-free graph languages. Another line of future work is to provide support for more general context-free graph languages using vertex replacement grammars. This would allow us to reason about more interesting families of diagrams and borrow proof techniques from the rich literature on context-free graph grammars.

\medskip

\noindent \textbf{Acknowledgements.} In addition to the two authors, Quantomatic has received major contributions from Alex Merry, Lucas Dixon, and Ross Duncan. We would also like to thank David Quick, Benjamin Frot, Fabio Zennaro, Krzysztof Bar, Gudmund Grov, Yuhui Lin, Matvey Soloviev, Song Zhang, and Michael Bradley for their contributions and gratefully acknowledge financial support from EPSRC, the Scatcherd European Scholarship, and the John Templeton Foundation.

\bibliographystyle{plain}
\bibliography{main,all}

\end{document}

%% file: figures/compound-process-new.tikz
\begin{tikzpicture}[scale=1.3]
	\begin{pgfonlayer}{nodelayer}
		\node [style=right label, yshift=0.5mm] (0) at (1, 2) {$C$};
		\node [style=none] (1) at (0.75, 1.5) {};
		\node [style=none] (2) at (1.25, -2.75) {};
		\node [style=right label, yshift=0.5mm] (3) at (-2, 2) {$A$};
		\node [style=right label] (4) at (1.25, 0) {$D$};
		\node [style=none] (5) at (1.25, -1.75) {};
		\node [style=none] (6) at (-0.75, 0.5) {};
		\node [style=right label, yshift=0.5mm] (7) at (-0.5, 2) {$B$};
		\node [style=none] (8) at (-0.75, 2.5) {};
		\node [style=semilarge box, yshift=0.2mm] (9) at (0, 1) {$g$};
		\node [style=none] (10) at (-2.25, 2.5) {};
		\node [style=right label] (11) at (1.5, -2.5) {$A$};
		\node [style=none] (12) at (-1.25, -0.75) {};
		\node [style=none] (13) at (0.75, 0.5) {};
		\node [style=medium box] (14) at (-1.75, -1.25) {$f$};
		\node [style=small box, yshift=0.2mm] (15) at (1.25, -1.25) {$h$};
		\node [style=right label] (16) at (-0.75, -0.25) {$A$};
		\node [style=none] (17) at (-2.25, -0.75) {};
		\node [style=none] (18) at (-0.75, 1.5) {};
		\node [style=none] (19) at (1.25, -0.75) {};
		\node [style=none] (20) at (0.75, 2.5) {};
	\end{pgfonlayer}
	\begin{pgfonlayer}{edgelayer}
		\draw [style=diredge, in=-90, out=90, looseness=1.00] (12.center) to (6.center);
		\draw [style=diredge] (18.center) to (8.center);
		\draw [style=diredge, in=-90, out=90, looseness=1.00] (19.center) to (13.center);
		\draw [style=diredge] (17.center) to (10.center);
		\draw [style=diredge] (2.center) to (5.center);
		\draw [style=diredge] (1.center) to (20.center);
	\end{pgfonlayer}
\end{tikzpicture}

%% file: figures/assoc_w_points.tikz
\begin{tikzpicture}[dotpic]
	\path [use as bounding box] (-3,-1.5) rectangle (3,1.5);
	\begin{pgfonlayer}{nodelayer}
		\node [style=white dot] (0) at (-2.5, -0.25) {};
		\node [style=white dot] (1) at (-2, 0.5) {};
		\node [style=none, yshift=0.2 mm] (2) at (-3, -1) {};
		\node [style=none] (3) at (-2, 1.5) {};
		\node [style=none] (4) at (0, 0) {$=$};
		\node [style=label] (5) at (-3, -1.5) {$a$};
		\node [style=label] (6) at (-2, -1.5) {$b$};
		\node [style=label] (7) at (-1, -1.5) {$c$};
		\node [style=label] (8) at (2, -1.5) {$b$};
		\node [style=label] (9) at (3, -1.5) {$c$};
		\node [style=label] (10) at (1, -1.5) {$a$};
		\node [style=none, yshift=0.2 mm] (11) at (-2, -1) {};
		\node [style=none, yshift=0.2 mm] (12) at (-1, -1) {};
		\node [style=white dot] (13) at (2.5, -0.25) {};
		\node [style=white dot] (14) at (2, 0.5) {};
		\node [style=none] (15) at (2, 1.5) {};
		\node [style=none, yshift=0.2 mm] (16) at (1, -1) {};
		\node [style=none, yshift=0.2 mm] (17) at (2, -1) {};
		\node [style=none, yshift=0.2 mm] (18) at (3, -1) {};
	\end{pgfonlayer}
	\begin{pgfonlayer}{edgelayer}
		\draw [style=diredge] (1) to (3.center);
		\draw [style=diredge] (0) to (1);
		\draw [style=diredge] (2.center) to (0);
		\draw [style=diredge] (11.center) to (0);
		\draw [style=diredge] (12.center) to (1);
		\draw [style=diredge] (14) to (15.center);
		\draw [style=diredge] (13) to (14);
		\draw [style=diredge] (18.center) to (13);
		\draw [style=diredge] (17.center) to (13);
		\draw [style=diredge] (16.center) to (14);
	\end{pgfonlayer}
\end{tikzpicture}

%% file: figures/unit_w_points.tikz
\begin{tikzpicture}[dotpic]
	\path [use as bounding box] (-3.25,-1.25) rectangle (3.25,1);
	\begin{pgfonlayer}{nodelayer}
		\node [style=white dot] (0) at (-2.75, 0) {};
		\node [style=none] (1) at (-2.75, 1) {};
		\node [style=white dot] (2) at (-2.25, -0.75) {};
		\node [style=none] (3) at (-1, 0) {$=$};
		\node [style=none] (4) at (0, 1) {};
		\node [style=white dot] (5) at (2.25, -0.75) {};
		\node [style=white dot] (6) at (2.75, 0) {};
		\node [style=none] (7) at (2.75, 1) {};
		\node [style=none] (8) at (1, 0) {$=$};
		\node [style=label] (9) at (-3.25, -1.25) {$a$};
		\node [style=label] (10) at (3.25, -1.25) {$a$};
		\node [style=label] (11) at (0, -1.25) {$a$};
		\node [style=none, yshift=0.2 mm] (12) at (3.25, -0.75) {};
		\node [style=none, yshift=0.2 mm] (13) at (-3.25, -0.75) {};
		\node [style=none, yshift=0.2 mm] (14) at (0, -0.75) {};
	\end{pgfonlayer}
	\begin{pgfonlayer}{edgelayer}
		\draw [style=diredge] (0) to (1.center);
		\draw [style=diredge] (2) to (0);
		\draw [style=diredge] (6) to (7.center);
		\draw [style=diredge] (5) to (6);
		\draw [style=diredge] (12.center) to (6);
		\draw [style=diredge] (13.center) to (0);
		\draw [style=diredge] (14.center) to (4.center);
	\end{pgfonlayer}
\end{tikzpicture}

%% file: figures/comm_w_points.tikz
\begin{tikzpicture}
	\path [use as bounding box] (-2.75,-1.25) rectangle (2.75,1.25);
	\begin{pgfonlayer}{nodelayer}
		\node [style=white dot] (0) at (-1.75, 0.25) {};
		\node [style=none] (1) at (-1.75, 1.25) {};
		\node [style=none, yshift=0.2mm] (2) at (-2.25, -0.75) {};
		\node [style=none, yshift=0.2 mm] (3) at (-1.25, -0.75) {};
		\node [style=none] (4) at (0, 0) {$=$};
		\node [style=label] (5) at (2.25, -1.25) {$b$};
		\node [style=label] (6) at (1.25, -1.25) {$a$};
		\node [style=none] (7) at (1.25, 0) {};
		\node [style=none, yshift=0.2 mm] (8) at (2.25, -0.75) {};
		\node [style=none] (9) at (2.25, 0) {};
		\node [style=none] (10) at (1.75, 1.25) {};
		\node [style=none, yshift=0.2 mm] (11) at (1.25, -0.75) {};
		\node [style=white dot] (12) at (1.75, 0.25) {};
		\node [style=label] (13) at (-1.25, -1.25) {$b$};
		\node [style=label] (14) at (-2.25, -1.25) {$a$};
	\end{pgfonlayer}
	\begin{pgfonlayer}{edgelayer}
		\draw [style=diredge] (0) to (1.center);
		\draw [style=diredge] (3.center) to (0);
		\draw [style=diredge] (2.center) to (0);
		\draw [style=diredge] (12) to (10);
		\draw [in=-75, out=60, looseness=0.75] (11.center) to (9.center);
		\draw [in=-105, out=120, looseness=0.75] (8.center) to (7.center);
		\draw [style=diredge, in=-165, out=75, looseness=1.00] (7.center) to (12);
		\draw [style=diredge, in=-15, out=105, looseness=1.00] (9.center) to (12);
	\end{pgfonlayer}
\end{tikzpicture}

%% file: figures/assoc_rewrite.tikz
\begin{tikzpicture}[dotpic]
	\begin{pgfonlayer}{nodelayer}
		\node [style=white dot] (0) at (-7, -0.5) {};
		\node [style=white dot] (1) at (-6.5, 0.25) {};
		\node [style=white dot] (2) at (-6.25, -2) {};
		\node [style=none] (3) at (-8, -2) {};
		\node [style=none] (4) at (-5, -2) {};
		\node [style=white dot] (5) at (-7.25, 1.75) {};
		\node [style=white dot] (6) at (-5.75, -2.75) {};
		\node [style=none] (7) at (-7.25, 2.5) {};
		\node [style=none] (8) at (-9.25, -1.25) {};
		\node [style=none] (9) at (-6.75, -3) {};
		\node [style=none] (10) at (-7.75, -1.25) {};
		\node [style=none] (11) at (-7.25, -1.25) {};
		\node [style=none] (12) at (-6.25, -1.5) {};
		\node [style=none] (13) at (-6.75, -1.5) {};
		\node [style=none] (14) at (-5.25, -1.25) {};
		\node [style=none] (15) at (-5.75, -1.25) {};
		\node [style=none] (16) at (-6.5, 0.75) {};
		\node [style=none] (17) at (-7, 0.75) {};
		\node [style=none] (18) at (0.25, -3) {};
		\node [style=none] (19) at (0.75, 0.75) {};
		\node [style=none] (20) at (2, -1.25) {};
		\node [style=none] (21) at (0.5, -1.5) {};
		\node [style=white dot] (22) at (0, 1.75) {};
		\node [style=none] (23) at (-2, -1.25) {};
		\node [style=none] (24) at (2.25, -2) {};
		\node [style=none] (25) at (0, 2.5) {};
		\node [style=none] (26) at (1, -1.5) {};
		\node [style=none] (27) at (1.5, -1.25) {};
		\node [style=none] (28) at (0, -1.25) {};
		\node [style=none] (29) at (-0.5, -1.25) {};
		\node [style=none] (30) at (0.25, 0.75) {};
		\node [style=white dot] (31) at (1.25, -2.75) {};
		\node [style=white dot] (32) at (0.75, -2) {};
		\node [style=none] (33) at (-0.75, -2) {};
		\node [style=none] (34) at (-0.25, -1.25) {};
		\node [style=none] (35) at (0.75, -1.5) {};
		\node [style=none] (36) at (1.75, -1.25) {};
		\node [style=none] (37) at (0.5, 0.75) {};
		\node [style=none] (38) at (-3.5, 0) {$\Rightarrow$};
		\node [style=white dot] (39) at (8, -2.75) {};
		\node [style=none] (40) at (8.5, -1.25) {};
		\node [style=none] (41) at (9, -1.25) {};
		\node [style=none] (42) at (7.25, 0.75) {};
		\node [style=white dot] (43) at (8.25, -0.5) {};
		\node [style=none] (44) at (9.25, -2) {};
		\node [style=none] (45) at (5, -1.25) {};
		\node [style=white dot] (46) at (7.75, 0.25) {};
		\node [style=none] (47) at (7, -3) {};
		\node [style=none] (48) at (7, -1.25) {};
		\node [style=white dot] (49) at (7, 1.75) {};
		\node [style=none] (50) at (8, -1.5) {};
		\node [style=none] (51) at (7.75, 0.75) {};
		\node [style=none] (52) at (7, 2.5) {};
		\node [style=white dot] (53) at (7.5, -2) {};
		\node [style=none] (54) at (6.25, -2) {};
		\node [style=none] (55) at (7.5, -1.5) {};
		\node [style=none] (56) at (6.5, -1.25) {};
		\node [style=none] (57) at (4, 0) {$\Rightarrow$};
	\end{pgfonlayer}
	\begin{pgfonlayer}{edgelayer}
		\draw [style=diredge] (1) to (5);
		\draw [style=diredge] (0) to (1);
		\draw [style=diredge] (3.center) to (0);
		\draw [style=diredge] (2) to (0);
		\draw [style=diredge] (4.center) to (1);
		\draw [style=diredge] (9.center) to (2);
		\draw [style=diredge] (6) to (2);
		\draw [style=diredge] (8.center) to (5);
		\draw [style=diredge] (5) to (7.center);
		\draw [style=dashed edge] (10.center) to (11.center);
		\draw [style=dashed edge] (13.center) to (12.center);
		\draw [style=dashed edge] (15.center) to (14.center);
		\draw [style=dashed edge] (17.center) to (16.center);
		\draw [style=diredge] (18.center) to (32);
		\draw [style=diredge] (31) to (32);
		\draw [style=diredge] (23.center) to (22);
		\draw [style=diredge] (22) to (25.center);
		\draw [style=dashed edge] (29.center) to (28.center);
		\draw [style=dashed edge] (21.center) to (26.center);
		\draw [style=dashed edge] (27.center) to (20.center);
		\draw [style=dashed edge] (30.center) to (19.center);
		\draw [style=diredge] (37.center) to (22);
		\draw (33.center) to (34.center);
		\draw (32) to (35.center);
		\draw (24.center) to (36.center);
		\draw [style=diredge] (46) to (49);
		\draw [style=diredge] (43) to (46);
		\draw [style=diredge] (47.center) to (53);
		\draw [style=diredge] (39) to (53);
		\draw [style=diredge] (45.center) to (49);
		\draw [style=diredge] (49) to (52.center);
		\draw [style=dashed edge] (56.center) to (48.center);
		\draw [style=dashed edge] (55.center) to (50.center);
		\draw [style=dashed edge] (40.center) to (41.center);
		\draw [style=dashed edge] (42.center) to (51.center);
		\draw [style=diredge] (54.center) to (46);
		\draw [style=diredge] (44.center) to (43);
		\draw [style=diredge] (53) to (43);
	\end{pgfonlayer}
\end{tikzpicture}

%% file: figures/coassoc.tikz
\begin{tikzpicture}[dotpic]
	\begin{pgfonlayer}{nodelayer}
		\node [style=gray dot] (0) at (-2, 0.75) {};
		\node [style=gray dot] (1) at (-1.5, 0) {};
		\node [style=none] (2) at (-2.5, 1.5) {};
		\node [style=none] (3) at (-1.5, 1.5) {};
		\node [style=none] (4) at (-0.5, 1.5) {};
		\node [style=none] (5) at (-1.5, -1) {};
		\node [style=none] (6) at (0, 0.5) {$=$};
		\node [style=none] (7) at (1.5, 1.5) {};
		\node [style=gray dot] (8) at (2, 0.75) {};
		\node [style=none] (9) at (0.5, 1.5) {};
		\node [style=gray dot] (10) at (1.5, 0) {};
		\node [style=none] (11) at (1.5, -1) {};
		\node [style=none] (12) at (2.5, 1.5) {};
	\end{pgfonlayer}
	\begin{pgfonlayer}{edgelayer}
		\draw [style=diredge] (5.center) to (1);
		\draw [style=diredge] (1) to (0);
		\draw [style=diredge] (0) to (3.center);
		\draw [style=diredge] (0) to (2.center);
		\draw [style=diredge] (1) to (4.center);
		\draw [style=diredge] (11.center) to (10);
		\draw [style=diredge] (10) to (8);
		\draw [style=diredge] (8) to (7.center);
		\draw [style=diredge] (8) to (12.center);
		\draw [style=diredge] (10) to (9.center);
	\end{pgfonlayer}
\end{tikzpicture}

%% file: figures/counit-law.tikz
\begin{tikzpicture}[dotpic]
	\begin{pgfonlayer}{nodelayer}
		\node [style=gray dot] (0) at (-2, 0) {};
		\node [style=none] (1) at (-2, -0.75) {};
		\node [style=none] (2) at (-2.5, 0.75) {};
		\node [style=gray dot] (3) at (-1.5, 0.75) {};
		\node [style=none] (4) at (-1, 0) {$=$};
		\node [style=none] (5) at (0, 0.75) {};
		\node [style=none] (6) at (0, -0.75) {};
		\node [style=none] (7) at (2.25, 0.75) {};
		\node [style=gray dot] (8) at (1.25, 0.75) {};
		\node [style=gray dot] (9) at (1.75, 0) {};
		\node [style=none] (10) at (1.75, -0.75) {};
		\node [style=none] (11) at (0.75, 0) {$=$};
	\end{pgfonlayer}
	\begin{pgfonlayer}{edgelayer}
		\draw [style=diredge] (1.center) to (0);
		\draw [style=diredge] (0) to (3);
		\draw [style=diredge] (0) to (2.center);
		\draw [style=diredge] (6.center) to (5.center);
		\draw [style=diredge] (10.center) to (9);
		\draw [style=diredge] (9) to (8);
		\draw [style=diredge] (9) to (7.center);
	\end{pgfonlayer}
\end{tikzpicture}

%% file: figures/cocomm.tikz
\begin{tikzpicture}
	\begin{pgfonlayer}{nodelayer}
		\node [style=none] (0) at (1.75, -1) {};
		\node [style=none] (1) at (1.25, 0.25) {};
		\node [style=none] (2) at (2.25, 0.25) {};
		\node [style=gray dot] (3) at (1.75, 0) {};
		\node [style=none] (4) at (2.25, 1) {};
		\node [style=none] (5) at (1.25, 1) {};
		\node [style=gray dot] (6) at (-1.75, 0) {};
		\node [style=none] (7) at (-1.75, -1) {};
		\node [style=none] (8) at (-2.25, 1) {};
		\node [style=none] (9) at (-1.25, 1) {};
		\node [style=none] (10) at (0, 0) {$=$};
	\end{pgfonlayer}
	\begin{pgfonlayer}{edgelayer}
		\draw [style=diredge] (0) to (3);
		\draw [style=diredge, in=-60, out=75, looseness=0.75] (2.center) to (5.center);
		\draw [style=diredge, in=-120, out=105, looseness=0.75] (1.center) to (4.center);
		\draw [in=-75, out=165, looseness=1.00] (3) to (1.center);
		\draw [in=-105, out=15, looseness=1.00] (3) to (2.center);
		\draw [style=diredge] (7) to (6);
		\draw [style=diredge] (6) to (9);
		\draw [style=diredge] (6) to (8);
	\end{pgfonlayer}
\end{tikzpicture}

%% file: figures/bialgebra.tikz
\begin{tikzpicture}[dotpic]
	\begin{pgfonlayer}{nodelayer}
		\node [style=none] (0) at (2, -1.25) {};
		\node [style=none] (1) at (1, -1.25) {};
		\node [style=none] (2) at (1, 1.25) {};
		\node [style=none] (3) at (2, 1.25) {};
		\node [style=gray dot] (4) at (2, -0.5) {};
		\node [style=gray dot] (5) at (1, -0.5) {};
		\node [style=white dot] (6) at (2, 0.5) {};
		\node [style=white dot] (7) at (1, 0.5) {};
		\node [style=none] (8) at (0, 0) {$=$};
		\node [style=white dot] (9) at (-1.5, -0.5) {};
		\node [style=gray dot] (10) at (-1.5, 0.5) {};
		\node [style=none] (11) at (-2.25, -1.25) {};
		\node [style=none] (12) at (-0.75, -1.25) {};
		\node [style=none] (13) at (-2.25, 1.25) {};
		\node [style=none] (14) at (-0.75, 1.25) {};
	\end{pgfonlayer}
	\begin{pgfonlayer}{edgelayer}
		\draw [style=diredge] (1.center) to (5);
		\draw [style=diredge] (5) to (6);
		\draw [style=diredge] (4) to (7);
		\draw [style=diredge] (5) to (7);
		\draw [style=diredge] (4) to (6);
		\draw [style=diredge] (0.center) to (4);
		\draw [style=diredge] (6) to (3.center);
		\draw [style=diredge] (7) to (2.center);
		\draw [style=diredge] (11.center) to (9);
		\draw [style=diredge] (12.center) to (9);
		\draw [style=diredge] (9) to (10);
		\draw [style=diredge] (10) to (13.center);
		\draw [style=diredge] (10) to (14.center);
	\end{pgfonlayer}
\end{tikzpicture}

%% file: figures/white-copy.tikz
\begin{tikzpicture}[dotpic]
	\begin{pgfonlayer}{nodelayer}
		\node [style=none] (0) at (0, 0) {$=$};
		\node [style=white dot] (1) at (-1.25, -0.75) {};
		\node [style=gray dot] (2) at (-1.25, 0) {};
		\node [style=none] (3) at (-1.75, 0.75) {};
		\node [style=none] (4) at (-0.75, 0.75) {};
		\node [style=none] (5) at (1, 0.75) {};
		\node [style=white dot] (6) at (1, -0.25) {};
		\node [style=white dot] (7) at (1.75, -0.25) {};
		\node [style=none] (8) at (1.75, 0.75) {};
	\end{pgfonlayer}
	\begin{pgfonlayer}{edgelayer}
		\draw [style=diredge] (1) to (2);
		\draw [style=diredge] (2) to (3.center);
		\draw [style=diredge] (2) to (4.center);
		\draw [style=diredge] (6) to (5.center);
		\draw [style=diredge] (7) to (8.center);
	\end{pgfonlayer}
\end{tikzpicture}

%% file: figures/gray-cocopy.tikz
\begin{tikzpicture}[dotpic]
	\begin{pgfonlayer}{nodelayer}
		\node [style=none] (0) at (0, 0) {$=$};
		\node [style=gray dot] (1) at (-1.25, 0.75) {};
		\node [style=white dot] (2) at (-1.25, 0) {};
		\node [style=none] (3) at (-1.75, -0.75) {};
		\node [style=none] (4) at (-0.75, -0.75) {};
		\node [style=none] (5) at (1, -0.75) {};
		\node [style=gray dot] (6) at (1, 0.25) {};
		\node [style=gray dot] (7) at (1.75, 0.25) {};
		\node [style=none] (8) at (1.75, -0.75) {};
	\end{pgfonlayer}
	\begin{pgfonlayer}{edgelayer}
		\draw [style=diredge] (2) to (1);
		\draw [style=diredge] (3.center) to (2);
		\draw [style=diredge] (4.center) to (2);
		\draw [style=diredge] (5.center) to (6);
		\draw [style=diredge] (8.center) to (7);
	\end{pgfonlayer}
\end{tikzpicture}

%% file: figures/bialg_app.tikz
\begin{tikzpicture}[dotpic]
	\begin{pgfonlayer}{nodelayer}
		\node [style=gray dot] (0) at (-5.75, 0.25) {};
		\node [style=white dot] (1) at (-5.75, -0.5) {};
		\node [style=none] (2) at (-2.5, 0) {$\Rightarrow$};
		\node [style=gray dot] (3) at (-5, -1.75) {};
		\node [style=white dot] (4) at (-5, 1.5) {};
		\node [style=none] (5) at (-5, -2.5) {};
		\node [style=none] (6) at (-5, 2.25) {};
		\node [style=none] (7) at (-5.5, 1) {};
		\node [style=none] (8) at (-5, 1) {};
		\node [style=none] (9) at (-5, -1.25) {};
		\node [style=none] (10) at (-5.5, -1.25) {};
		\node [style=none] (11) at (3.25, 0) {$\Rightarrow$};
		\node [style=none] (12) at (-6, -1.25) {};
		\node [style=none] (13) at (-6.5, -1.25) {};
		\node [style=none] (14) at (-6, 1) {};
		\node [style=none] (15) at (-6.5, 1) {};
		\node [style=none] (16) at (-6.25, -1.25) {};
		\node [style=none] (17) at (-6.25, 1) {};
		\node [style=none] (18) at (-6.25, 2.25) {};
		\node [style=none] (19) at (-6.25, -2.5) {};
		\node [style=none] (20) at (-5.25, -1.25) {};
		\node [style=none] (21) at (-5.25, 1) {};
		\node [style=none] (22) at (0.75, -1.25) {};
		\node [style=none] (23) at (-0.25, -1.25) {};
		\node [style=none] (24) at (-0.5, 2.25) {};
		\node [style=none] (25) at (0.75, -2.5) {};
		\node [style=none] (26) at (-0.75, 1) {};
		\node [style=gray dot] (27) at (0.75, -1.75) {};
		\node [style=none] (28) at (0.75, 1) {};
		\node [style=none] (29) at (-0.5, -2.5) {};
		\node [style=none] (30) at (-0.75, -1.25) {};
		\node [style=none] (31) at (-0.5, -1.25) {};
		\node [style=none] (32) at (0.25, -1.25) {};
		\node [style=none] (33) at (-0.5, 1) {};
		\node [style=white dot] (34) at (0.75, 1.5) {};
		\node [style=none] (35) at (0.25, 1) {};
		\node [style=none] (36) at (0.5, -1.25) {};
		\node [style=none] (37) at (0.75, 2.25) {};
		\node [style=none] (38) at (0.5, 1) {};
		\node [style=none] (39) at (-0.25, 1) {};
		\node [style=none] (40) at (5.75, -1.25) {};
		\node [style=none] (41) at (4.75, -1.25) {};
		\node [style=none] (42) at (4.5, 2.25) {};
		\node [style=none] (43) at (5.75, -2.5) {};
		\node [style=none] (44) at (4.25, 1) {};
		\node [style=gray dot] (45) at (5.75, -1.75) {};
		\node [style=none] (46) at (5.75, 1) {};
		\node [style=none] (47) at (4.5, -2.5) {};
		\node [style=none] (48) at (4.25, -1.25) {};
		\node [style=none] (49) at (4.5, -1.25) {};
		\node [style=none] (50) at (5.25, -1.25) {};
		\node [style=none] (51) at (4.5, 1) {};
		\node [style=white dot] (52) at (5.75, 1.5) {};
		\node [style=none] (53) at (5.25, 1) {};
		\node [style=none] (54) at (5.5, -1.25) {};
		\node [style=none] (55) at (5.75, 2.25) {};
		\node [style=none] (56) at (5.5, 1) {};
		\node [style=none] (57) at (4.75, 1) {};
		\node [style=gray dot] (58) at (4.5, -0.5) {};
		\node [style=gray dot] (59) at (5.5, -0.5) {};
		\node [style=white dot] (60) at (5.5, 0.5) {};
		\node [style=white dot] (61) at (4.5, 0.5) {};
	\end{pgfonlayer}
	\begin{pgfonlayer}{edgelayer}
		\draw [style=diredge] (5.center) to (3);
		\draw [style=diredge] (4) to (6.center);
		\draw [style=dashed edge] (7.center) to (8.center);
		\draw [style=dashed edge] (10.center) to (9.center);
		\draw [style=diredge, bend right=75, looseness=1.50] (3) to (4);
		\draw [style=diredge] (1) to (0);
		\draw [style=dashed edge] (13.center) to (12.center);
		\draw [style=dashed edge] (15.center) to (14.center);
		\draw [in=-90, out=135, looseness=1.25] (3) to (20.center);
		\draw (19.center) to (16.center);
		\draw [style=diredge, in=-135, out=90, looseness=1.25] (21.center) to (4);
		\draw [style=diredge] (17.center) to (18.center);
		\draw [style=diredge, in=-30, out=90, looseness=1.00] (20.center) to (1);
		\draw [style=diredge, in=-150, out=90, looseness=1.00] (16.center) to (1);
		\draw [in=-90, out=30, looseness=1.00] (0) to (21.center);
		\draw [in=-90, out=150, looseness=1.00] (0) to (17.center);
		\draw [style=diredge] (25.center) to (27);
		\draw [style=diredge] (34) to (37.center);
		\draw [style=dashed edge] (35.center) to (28.center);
		\draw [style=dashed edge] (32.center) to (22.center);
		\draw [style=diredge, bend right=75, looseness=1.50] (27) to (34);
		\draw [style=dashed edge] (30.center) to (23.center);
		\draw [style=dashed edge] (26.center) to (39.center);
		\draw [in=-90, out=135, looseness=1.25] (27) to (36.center);
		\draw (29.center) to (31.center);
		\draw [style=diredge, in=-135, out=90, looseness=1.25] (38.center) to (34);
		\draw [style=diredge] (33.center) to (24.center);
		\draw [style=diredge] (43.center) to (45);
		\draw [style=diredge] (52) to (55.center);
		\draw [style=dashed edge] (53.center) to (46.center);
		\draw [style=dashed edge] (50.center) to (40.center);
		\draw [style=diredge, bend right=75, looseness=1.50] (45) to (52);
		\draw [style=dashed edge] (48.center) to (41.center);
		\draw [style=dashed edge] (44.center) to (57.center);
		\draw [in=-90, out=135, looseness=1.25] (45) to (54.center);
		\draw (47.center) to (49.center);
		\draw [style=diredge, in=-135, out=90, looseness=1.25] (56.center) to (52);
		\draw [style=diredge] (51.center) to (42.center);
		\draw [style=diredge] (54.center) to (59);
		\draw [style=diredge] (49.center) to (58);
		\draw [style=diredge] (59) to (61);
		\draw [style=diredge] (58) to (60);
		\draw [style=diredge] (59) to (60);
		\draw [style=diredge] (58) to (61);
		\draw (60) to (56.center);
		\draw (61) to (51.center);
	\end{pgfonlayer}
\end{tikzpicture}

%% file: figures/nary-merge.tikz
\begin{tikzpicture}
	\begin{pgfonlayer}{nodelayer}
		\node [style=none] (0) at (0, 0) {$=$};
		\node [style=none] (1) at (-2.25, 1.5) {};
		\node [style=white dot] (2) at (-2.25, 0.5) {};
		\node [style=none] (3) at (-3.75, -1) {};
		\node [style=white dot] (4) at (-1.5, -0.25) {};
		\node [style=none] (5) at (-2, -1) {};
		\node [style=white dot] (6) at (1.75, 0) {};
		\node [style=none] (7) at (2.5, -1) {};
		\node [style=none] (8) at (1.75, 1.25) {};
		\node [style=none] (9) at (1, -1) {};
		\node [style=none] (10) at (-2.75, -1) {};
		\node [style=none] (11) at (-1, -1) {};
		\node [style=none] (12) at (-3.25, -1) {...};
		\node [style=none] (13) at (-1.5, -1) {...};
		\node [style=none] (14) at (1.75, -1) {...};
	\end{pgfonlayer}
	\begin{pgfonlayer}{edgelayer}
		\draw [style=diredge] (2) to (1.center);
		\draw [style=diredge] (4) to (2);
		\draw [style=diredge] (5.center) to (4);
		\draw [style=diredge] (9.center) to (6);
		\draw [style=diredge] (6) to (8.center);
		\draw [style=diredge] (7.center) to (6);
		\draw [style=diredge] (11.center) to (4);
		\draw [style=diredge] (10.center) to (2);
		\draw [style=diredge] (3.center) to (2);
	\end{pgfonlayer}
\end{tikzpicture}

%% file: figures/id-to-wire.tikz
\begin{tikzpicture}
	\begin{pgfonlayer}{nodelayer}
		\node [style=none] (0) at (0, 0) {$=$};
		\node [style=none] (1) at (-1.25, 1) {};
		\node [style=white dot] (2) at (-1.25, 0) {};
		\node [style=none] (3) at (1, 1) {};
		\node [style=none] (4) at (1, -1) {};
		\node [style=none] (5) at (-1.25, -1) {};
	\end{pgfonlayer}
	\begin{pgfonlayer}{edgelayer}
		\draw [style=diredge] (2) to (1.center);
		\draw [style=diredge] (5.center) to (2);
		\draw [style=diredge] (4.center) to (3.center);
	\end{pgfonlayer}
\end{tikzpicture}

%% file: figures/bb-tree-merge.tikz
\begin{tikzpicture}
	\begin{pgfonlayer}{nodelayer}
		\node [style=none] (0) at (0, 0) {$=$};
		\node [style=none] (1) at (-3, -2) {};
		\node [style=none] (2) at (-4, -2) {};
		\node [style=none] (3) at (-3, -1) {};
		\node [style=bbox corner] (4) at (-4, -1) {};
		\node [style=none] (5) at (-2.25, 1.75) {};
		\node [style=white dot] (6) at (-2.25, 0.75) {};
		\node [style=none] (7) at (-3.5, -1.5) {};
		\node [style=none, blue] (8) at (-4.5, -1) {$b$};
		\node [style=white dot] (9) at (-1.5, 0) {};
		\node [style=none, blue] (10) at (-2.5, -1) {$c$};
		\node [style=none] (11) at (-2, -2) {};
		\node [style=bbox corner] (12) at (-2, -1) {};
		\node [style=none] (13) at (-1.5, -1.5) {};
		\node [style=none] (14) at (-1, -2) {};
		\node [style=none] (15) at (-1, -1) {};
		\node [style=none] (16) at (2, -2) {};
		\node [style=none, blue] (17) at (2.5, -1) {$c$};
		\node [style=white dot] (18) at (2.5, 0.25) {};
		\node [style=none] (19) at (3.5, -1.5) {};
		\node [style=none] (20) at (2.5, 1.75) {};
		\node [style=bbox corner] (21) at (3, -1) {};
		\node [style=none] (22) at (2, -1) {};
		\node [style=none] (23) at (3, -2) {};
		\node [style=none] (24) at (1.5, -1.5) {};
		\node [style=none] (25) at (1, -2) {};
		\node [style=none] (26) at (4, -1) {};
		\node [style=none] (27) at (4, -2) {};
		\node [style=bbox corner] (28) at (1, -1) {};
		\node [style=none, blue] (29) at (0.5, -1) {$b$};
	\end{pgfonlayer}
	\begin{pgfonlayer}{edgelayer}
		\draw [style=diredge] (7.center) to (6);
		\draw [style=diredge] (6) to (5.center);
		\draw [style=bbox edge] (4) to (3.center);
		\draw [style=bbox edge] (3.center) to (1.center);
		\draw [style=bbox edge] (1.center) to (2.center);
		\draw [style=bbox edge] (2.center) to (4);
		\draw [style=diredge] (9) to (6);
		\draw [style=bbox edge] (12) to (15.center);
		\draw [style=bbox edge] (15.center) to (14.center);
		\draw [style=bbox edge] (14.center) to (11.center);
		\draw [style=bbox edge] (11.center) to (12);
		\draw [style=diredge] (13.center) to (9);
		\draw [style=diredge] (24.center) to (18);
		\draw [style=diredge] (18) to (20.center);
		\draw [style=bbox edge] (28) to (22.center);
		\draw [style=bbox edge] (22.center) to (16.center);
		\draw [style=bbox edge] (16.center) to (25.center);
		\draw [style=bbox edge] (25.center) to (28);
		\draw [style=bbox edge] (21) to (26.center);
		\draw [style=bbox edge] (26.center) to (27.center);
		\draw [style=bbox edge] (27.center) to (23.center);
		\draw [style=bbox edge] (23.center) to (21);
		\draw [style=diredge] (19.center) to (18);
	\end{pgfonlayer}
\end{tikzpicture}

%% file: figures/quanto-output-thm.tikz
\begin{tikzpicture}[baseline={([yshift=-.5ex]current bounding box.center)}]
	\begin{pgfonlayer}{nodelayer}
		\node [style=wire] (b5) at (-2.75, -4.0) {};
		\node [style=X] (v4) at (0.5, -2.25) {};
		\node [style=wire] (b1) at (-1.25, 4.0) {};
		\node [style=wire] (b0) at (-2.5, 4.0) {};
		\node [style=X] (v3) at (-2.75, -2.25) {};
		\node [style=wire] (b2) at (0.75, 4.0) {};
		\node [style=Z] (v1) at (0.5, -3.5) {};
		\node [style=X] (v0) at (-1.0, 1.0) {};
		\node [style=wire] (b7) at (3.5, -4.0) {};
		\node [style=Z] (v7) at (-1.0, -0.25) {};
		\node [style=X] (v2) at (-0.25, 2.5) {};
		\node [style=Z] (v5) at (-4.0, 0.0) {};
		\node [style=wire] (b4) at (-5.0, -4.0) {};
		\node [style=wire] (b3) at (2.0, 4.0) {};
		\node [style=Z] (v8) at (2.0, 0.0) {};
		\node [style=Z] (v6) at (-2.5, 2.5) {};
		\node [style=none] (padl) at (-6.0, -5.0) {};
		\node [style=none] (padr) at (4.5, 5.0) {};
		\node [style=none] (padu) at (-6.0, 5.0) {};
		\node [style=none] (padd) at (4.5, -5.0) {};
	\end{pgfonlayer}
	\begin{pgfonlayer}{edgelayer}
		\draw [style=directed] (v8) to (b3);
		\draw [style=directed] (b4) to (v5);
		\draw [style=directed] (v5) to (v6);
		\draw [style=directed] (b7) to (v8);
		\draw [style=directed] (v0) to (v6);
		\draw [style=directed] (v0) to (v2);
		\draw [style=directed] (v7) to (v0);
		\draw [style=directed] (v1) to (v4);
		\draw [style=directed] (v2) to (b2);
		\draw [style=directed] (v3) to (v5);
		\draw [style=directed] (v3) to (v7);
		\draw [style=directed] (v6) to (b0);
		\draw [style=directed] (v2) to (b1);
		\draw [style=directed] (v4) to (v8);
		\draw [style=directed] (v4) to (v7);
		\draw [style=directed] (b5) to (v3);
	\end{pgfonlayer}
\end{tikzpicture}
=
\begin{tikzpicture}[baseline={([yshift=-.5ex]current bounding box.center)}]
	\begin{pgfonlayer}{nodelayer}
		\node [style=wire] (b5) at (-2.75, -4.0) {};
		\node [style=wire] (b1) at (-1.25, 4.0) {};
		\node [style=wire] (b0) at (-2.5, 4.0) {};
		\node [style=wire] (b2) at (0.75, 4.0) {};
		\node [style=Z] (v1) at (-3.0, 1.5) {};
		\node [style=X] (v0) at (-1.64, -0.29) {};
		\node [style=wire] (b7) at (3.5, -4.0) {};
		\node [style=wire] (b4) at (-5.0, -4.0) {};
		\node [style=wire] (b3) at (2.0, 4.0) {};
		\node [style=none] (padl) at (-6.0, -5.0) {};
		\node [style=none] (padr) at (4.5, 5.0) {};
		\node [style=none] (padu) at (-6.0, 5.0) {};
		\node [style=none] (padd) at (4.5, -5.0) {};
	\end{pgfonlayer}
	\begin{pgfonlayer}{edgelayer}
		\draw [style=directed] (b7) to (b3);
		\draw [style=directed] (b4) to (v1);
		\draw [style=directed, bend left=50.0] (v0) to (v1);
		\draw [style=directed, bend right=50.0] (v0) to (v1);
		\draw [style=directed] (v0) to (b2);
		\draw [style=directed] (v1) to (b0);
		\draw [style=directed] (v0) to (b1);
		\draw [style=directed] (b5) to (v0);
	\end{pgfonlayer}
\end{tikzpicture}

%% file: figures/quanto-output.tikz
\begin{quote}\raggedright
\begin{tikzpicture}[baseline={([yshift=-.5ex]current bounding box.center)}]
	\begin{pgfonlayer}{nodelayer}
		\node [style=wire] (b5) at (-2.75, -4.0) {};
		\node [style=X] (v4) at (0.5, -2.25) {};
		\node [style=wire] (b1) at (-1.25, 4.0) {};
		\node [style=wire] (b0) at (-2.5, 4.0) {};
		\node [style=X] (v3) at (-2.75, -2.25) {};
		\node [style=wire] (b2) at (0.75, 4.0) {};
		\node [style=Z] (v1) at (0.5, -3.5) {};
		\node [style=X] (v0) at (-1.0, 1.0) {};
		\node [style=wire] (b7) at (3.5, -4.0) {};
		\node [style=Z] (v7) at (-1.0, -0.25) {};
		\node [style=X] (v2) at (-0.25, 2.5) {};
		\node [style=Z] (v5) at (-4.0, 0.0) {};
		\node [style=wire] (b4) at (-5.0, -4.0) {};
		\node [style=wire] (b3) at (2.0, 4.0) {};
		\node [style=Z] (v8) at (2.0, 0.0) {};
		\node [style=Z] (v6) at (-2.5, 2.5) {};
		\node [style=none] (padl) at (-6.0, -5.0) {};
		\node [style=none] (padr) at (4.5, 5.0) {};
		\node [style=none] (padu) at (-6.0, 5.0) {};
		\node [style=none] (padd) at (4.5, -5.0) {};
	\end{pgfonlayer}
	\begin{pgfonlayer}{edgelayer}
		\draw [style=directed] (v8) to (b3);
		\draw [style=directed] (b4) to (v5);
		\draw [style=directed] (v5) to (v6);
		\draw [style=directed] (b7) to (v8);
		\draw [style=directed] (v0) to (v6);
		\draw [style=directed] (v0) to (v2);
		\draw [style=directed] (v7) to (v0);
		\draw [style=directed] (v1) to (v4);
		\draw [style=directed] (v2) to (b2);
		\draw [style=directed] (v3) to (v5);
		\draw [style=directed] (v3) to (v7);
		\draw [style=directed] (v6) to (b0);
		\draw [style=directed] (v2) to (b1);
		\draw [style=directed] (v4) to (v8);
		\draw [style=directed] (v4) to (v7);
		\draw [style=directed] (b5) to (v3);
	\end{pgfonlayer}
\end{tikzpicture}
=
\begin{tikzpicture}[baseline={([yshift=-.5ex]current bounding box.center)}]
	\begin{pgfonlayer}{nodelayer}
		\node [style=wire] (b5) at (-2.75, -4.0) {};
		\node [style=wire] (b1) at (-1.25, 4.0) {};
		\node [style=wire] (b0) at (-2.5, 4.0) {};
		\node [style=X] (v3) at (-2.75, -2.25) {};
		\node [style=Z] (v11) at (1.5, -2.25) {};
		\node [style=wire] (b2) at (0.75, 4.0) {};
		\node [style=Z] (v10) at (0.0, -2.25) {};
		\node [style=X] (v0) at (-1.0, 1.0) {};
		\node [style=wire] (b7) at (3.5, -4.0) {};
		\node [style=Z] (v7) at (-1.0, -0.25) {};
		\node [style=X] (v2) at (-0.25, 2.5) {};
		\node [style=Z] (v5) at (-4.0, 0.0) {};
		\node [style=wire] (b4) at (-5.0, -4.0) {};
		\node [style=wire] (b3) at (2.0, 4.0) {};
		\node [style=Z] (v8) at (2.5, 0.0) {};
		\node [style=Z] (v6) at (-2.5, 2.5) {};
		\node [style=none] (padl) at (-6.0, -5.0) {};
		\node [style=none] (padr) at (4.5, 5.0) {};
		\node [style=none] (padu) at (-6.0, 5.0) {};
		\node [style=none] (padd) at (4.5, -5.0) {};
	\end{pgfonlayer}
	\begin{pgfonlayer}{edgelayer}
		\draw [style=directed] (v11) to (v8);
		\draw [style=directed] (v8) to (b3);
		\draw [style=directed] (b4) to (v5);
		\draw [style=directed] (v5) to (v6);
		\draw [style=directed] (v10) to (v7);
		\draw [style=directed] (b7) to (v8);
		\draw [style=directed] (v0) to (v6);
		\draw [style=directed] (v0) to (v2);
		\draw [style=directed] (v7) to (v0);
		\draw [style=directed] (v2) to (b2);
		\draw [style=directed] (v3) to (v5);
		\draw [style=directed] (v3) to (v7);
		\draw [style=directed] (v6) to (b0);
		\draw [style=directed] (v2) to (b1);
		\draw [style=directed] (b5) to (v3);
	\end{pgfonlayer}
\end{tikzpicture}
=
\begin{tikzpicture}[baseline={([yshift=-.5ex]current bounding box.center)}]
	\begin{pgfonlayer}{nodelayer}
		\node [style=wire] (b5) at (-2.75, -4.0) {};
		\node [style=wire] (b1) at (-1.25, 4.0) {};
		\node [style=wire] (b0) at (-2.5, 4.0) {};
		\node [style=X] (v3) at (-2.75, -2.25) {};
		\node [style=wire] (b2) at (0.75, 4.0) {};
		\node [style=Z] (v1) at (2.75, 0.5) {};
		\node [style=Z] (v10) at (0.5, -2.25) {};
		\node [style=X] (v0) at (-1.0, 1.0) {};
		\node [style=wire] (b7) at (3.5, -4.0) {};
		\node [style=Z] (v7) at (-1.0, -0.25) {};
		\node [style=X] (v2) at (-0.25, 2.5) {};
		\node [style=Z] (v5) at (-4.0, 0.0) {};
		\node [style=wire] (b4) at (-5.0, -4.0) {};
		\node [style=wire] (b3) at (2.0, 4.0) {};
		\node [style=Z] (v6) at (-2.5, 2.5) {};
		\node [style=none] (padl) at (-6.0, -5.0) {};
		\node [style=none] (padr) at (4.5, 5.0) {};
		\node [style=none] (padu) at (-6.0, 5.0) {};
		\node [style=none] (padd) at (4.5, -5.0) {};
	\end{pgfonlayer}
	\begin{pgfonlayer}{edgelayer}
		\draw [style=directed] (v1) to (b3);
		\draw [style=directed] (b4) to (v5);
		\draw [style=directed] (v5) to (v6);
		\draw [style=directed] (v10) to (v7);
		\draw [style=directed] (b7) to (v1);
		\draw [style=directed] (v0) to (v6);
		\draw [style=directed] (v0) to (v2);
		\draw [style=directed] (v7) to (v0);
		\draw [style=directed] (v2) to (b2);
		\draw [style=directed] (v3) to (v5);
		\draw [style=directed] (v3) to (v7);
		\draw [style=directed] (v6) to (b0);
		\draw [style=directed] (v2) to (b1);
		\draw [style=directed] (b5) to (v3);
	\end{pgfonlayer}
\end{tikzpicture}
=
\begin{tikzpicture}[baseline={([yshift=-.5ex]current bounding box.center)}]
	\begin{pgfonlayer}{nodelayer}
		\node [style=wire] (b5) at (-2.75, -4.0) {};
		\node [style=Z] (v4) at (-1.75, -0.5) {};
		\node [style=wire] (b1) at (-1.25, 4.0) {};
		\node [style=wire] (b0) at (-2.5, 4.0) {};
		\node [style=X] (v3) at (-2.75, -2.25) {};
		\node [style=wire] (b2) at (0.75, 4.0) {};
		\node [style=Z] (v1) at (2.75, 0.5) {};
		\node [style=X] (v0) at (-1.0, 1.0) {};
		\node [style=wire] (b7) at (3.5, -4.0) {};
		\node [style=X] (v2) at (-0.25, 2.5) {};
		\node [style=Z] (v5) at (-4.0, 0.0) {};
		\node [style=wire] (b4) at (-5.0, -4.0) {};
		\node [style=wire] (b3) at (2.0, 4.0) {};
		\node [style=Z] (v6) at (-2.5, 2.5) {};
		\node [style=none] (padl) at (-6.0, -5.0) {};
		\node [style=none] (padr) at (4.5, 5.0) {};
		\node [style=none] (padu) at (-6.0, 5.0) {};
		\node [style=none] (padd) at (4.5, -5.0) {};
	\end{pgfonlayer}
	\begin{pgfonlayer}{edgelayer}
		\draw [style=directed] (v1) to (b3);
		\draw [style=directed] (b4) to (v5);
		\draw [style=directed] (v5) to (v6);
		\draw [style=directed] (b7) to (v1);
		\draw [style=directed] (v0) to (v6);
		\draw [style=directed] (v0) to (v2);
		\draw [style=directed] (v4) to (v0);
		\draw [style=directed] (v2) to (b2);
		\draw [style=directed] (v3) to (v5);
		\draw [style=directed] (v3) to (v4);
		\draw [style=directed] (v6) to (b0);
		\draw [style=directed] (v2) to (b1);
		\draw [style=directed] (b5) to (v3);
	\end{pgfonlayer}
\end{tikzpicture}
=
\begin{tikzpicture}[baseline={([yshift=-.5ex]current bounding box.center)}]
	\begin{pgfonlayer}{nodelayer}
		\node [style=wire] (b5) at (-2.75, -4.0) {};
		\node [style=wire] (b1) at (-1.25, 4.0) {};
		\node [style=wire] (b0) at (-2.5, 4.0) {};
		\node [style=X] (v3) at (-2.75, -2.25) {};
		\node [style=wire] (b2) at (0.75, 4.0) {};
		\node [style=Z] (v1) at (2.75, 0.5) {};
		\node [style=X] (v0) at (-1.0, 1.0) {};
		\node [style=wire] (b7) at (3.5, -4.0) {};
		\node [style=X] (v2) at (-0.25, 2.5) {};
		\node [style=Z] (v5) at (-4.0, 0.0) {};
		\node [style=wire] (b4) at (-5.0, -4.0) {};
		\node [style=wire] (b3) at (2.0, 4.0) {};
		\node [style=Z] (v6) at (-2.5, 2.5) {};
		\node [style=none] (padl) at (-6.0, -5.0) {};
		\node [style=none] (padr) at (4.5, 5.0) {};
		\node [style=none] (padu) at (-6.0, 5.0) {};
		\node [style=none] (padd) at (4.5, -5.0) {};
	\end{pgfonlayer}
	\begin{pgfonlayer}{edgelayer}
		\draw [style=directed] (v1) to (b3);
		\draw [style=directed] (b4) to (v5);
		\draw [style=directed] (v5) to (v6);
		\draw [style=directed] (b7) to (v1);
		\draw [style=directed] (v0) to (v6);
		\draw [style=directed] (v0) to (v2);
		\draw [style=directed] (v3) to (v0);
		\draw [style=directed] (v2) to (b2);
		\draw [style=directed] (v3) to (v5);
		\draw [style=directed] (v6) to (b0);
		\draw [style=directed] (v2) to (b1);
		\draw [style=directed] (b5) to (v3);
	\end{pgfonlayer}
\end{tikzpicture}
=
\begin{tikzpicture}[baseline={([yshift=-.5ex]current bounding box.center)}]
	\begin{pgfonlayer}{nodelayer}
		\node [style=wire] (b5) at (-2.75, -4.0) {};
		\node [style=wire] (b1) at (-1.25, 4.0) {};
		\node [style=wire] (b0) at (-2.5, 4.0) {};
		\node [style=X] (v3) at (-2.75, -2.25) {};
		\node [style=wire] (b2) at (0.75, 4.0) {};
		\node [style=X] (v0) at (-1.0, 1.0) {};
		\node [style=wire] (b7) at (3.5, -4.0) {};
		\node [style=X] (v2) at (-0.25, 2.5) {};
		\node [style=Z] (v5) at (-4.0, 0.0) {};
		\node [style=wire] (b4) at (-5.0, -4.0) {};
		\node [style=wire] (b3) at (2.0, 4.0) {};
		\node [style=Z] (v6) at (-2.5, 2.5) {};
		\node [style=none] (padl) at (-6.0, -5.0) {};
		\node [style=none] (padr) at (4.5, 5.0) {};
		\node [style=none] (padu) at (-6.0, 5.0) {};
		\node [style=none] (padd) at (4.5, -5.0) {};
	\end{pgfonlayer}
	\begin{pgfonlayer}{edgelayer}
		\draw [style=directed] (b7) to (b3);
		\draw [style=directed] (b4) to (v5);
		\draw [style=directed] (v5) to (v6);
		\draw [style=directed] (v0) to (v6);
		\draw [style=directed] (v0) to (v2);
		\draw [style=directed] (v3) to (v0);
		\draw [style=directed] (v2) to (b2);
		\draw [style=directed] (v3) to (v5);
		\draw [style=directed] (v6) to (b0);
		\draw [style=directed] (v2) to (b1);
		\draw [style=directed] (b5) to (v3);
	\end{pgfonlayer}
\end{tikzpicture}
=
\begin{tikzpicture}[baseline={([yshift=-.5ex]current bounding box.center)}]
	\begin{pgfonlayer}{nodelayer}
		\node [style=wire] (b5) at (-2.75, -4.0) {};
		\node [style=wire] (b1) at (-1.25, 4.0) {};
		\node [style=wire] (b0) at (-2.5, 4.0) {};
		\node [style=X] (v3) at (-2.75, -2.25) {};
		\node [style=wire] (b2) at (0.75, 4.0) {};
		\node [style=Z] (v1) at (-3.0, 1.5) {};
		\node [style=X] (v0) at (-1.5, -0.25) {};
		\node [style=wire] (b7) at (3.5, -4.0) {};
		\node [style=X] (v2) at (-0.25, 2.5) {};
		\node [style=wire] (b4) at (-5.0, -4.0) {};
		\node [style=wire] (b3) at (2.0, 4.0) {};
		\node [style=none] (padl) at (-6.0, -5.0) {};
		\node [style=none] (padr) at (4.5, 5.0) {};
		\node [style=none] (padu) at (-6.0, 5.0) {};
		\node [style=none] (padd) at (4.5, -5.0) {};
	\end{pgfonlayer}
	\begin{pgfonlayer}{edgelayer}
		\draw [style=directed] (b7) to (b3);
		\draw [style=directed] (b4) to (v1);
		\draw [style=directed] (v0) to (v2);
		\draw [style=directed] (v0) to (v1);
		\draw [style=directed] (v3) to (v0);
		\draw [style=directed] (v3) to (v1);
		\draw [style=directed] (v2) to (b2);
		\draw [style=directed] (v1) to (b0);
		\draw [style=directed] (v2) to (b1);
		\draw [style=directed] (b5) to (v3);
	\end{pgfonlayer}
\end{tikzpicture}
=
\begin{tikzpicture}[baseline={([yshift=-.5ex]current bounding box.center)}]
	\begin{pgfonlayer}{nodelayer}
		\node [style=wire] (b5) at (-2.75, -4.0) {};
		\node [style=X] (v4) at (-2.0, -0.25) {};
		\node [style=wire] (b1) at (-1.25, 4.0) {};
		\node [style=wire] (b0) at (-2.5, 4.0) {};
		\node [style=wire] (b2) at (0.75, 4.0) {};
		\node [style=Z] (v1) at (-3.0, 1.5) {};
		\node [style=wire] (b7) at (3.5, -4.0) {};
		\node [style=X] (v2) at (-0.25, 2.5) {};
		\node [style=wire] (b4) at (-5.0, -4.0) {};
		\node [style=wire] (b3) at (2.0, 4.0) {};
		\node [style=none] (padl) at (-6.0, -5.0) {};
		\node [style=none] (padr) at (4.5, 5.0) {};
		\node [style=none] (padu) at (-6.0, 5.0) {};
		\node [style=none] (padd) at (4.5, -5.0) {};
	\end{pgfonlayer}
	\begin{pgfonlayer}{edgelayer}
		\draw [style=directed] (b7) to (b3);
		\draw [style=directed] (b4) to (v1);
		\draw [style=directed] (v4) to (v2);
		\draw [style=directed, bend left=50.0] (v4) to (v1);
		\draw [style=directed, bend right=50.0] (v4) to (v1);
		\draw [style=directed] (v2) to (b2);
		\draw [style=directed] (v1) to (b0);
		\draw [style=directed] (v2) to (b1);
		\draw [style=directed] (b5) to (v4);
	\end{pgfonlayer}
\end{tikzpicture}
=
\begin{tikzpicture}[baseline={([yshift=-.5ex]current bounding box.center)}]
	\begin{pgfonlayer}{nodelayer}
		\node [style=wire] (b5) at (-2.75, -4.0) {};
		\node [style=wire] (b1) at (-1.25, 4.0) {};
		\node [style=wire] (b0) at (-2.5, 4.0) {};
		\node [style=wire] (b2) at (0.75, 4.0) {};
		\node [style=Z] (v1) at (-3.0, 1.5) {};
		\node [style=X] (v0) at (-1.6333333333333333, -0.2833333333333333) {};
		\node [style=wire] (b7) at (3.5, -4.0) {};
		\node [style=wire] (b4) at (-5.0, -4.0) {};
		\node [style=wire] (b3) at (2.0, 4.0) {};
		\node [style=none] (padl) at (-6.0, -5.0) {};
		\node [style=none] (padr) at (4.5, 5.0) {};
		\node [style=none] (padu) at (-6.0, 5.0) {};
		\node [style=none] (padd) at (4.5, -5.0) {};
	\end{pgfonlayer}
	\begin{pgfonlayer}{edgelayer}
		\draw [style=directed] (b7) to (b3);
		\draw [style=directed] (b4) to (v1);
		\draw [style=directed, bend left=50.0] (v0) to (v1);
		\draw [style=directed, bend right=50.0] (v0) to (v1);
		\draw [style=directed] (v0) to (b2);
		\draw [style=directed] (v1) to (b0);
		\draw [style=directed] (v0) to (b1);
		\draw [style=directed] (b5) to (v0);
	\end{pgfonlayer}
\end{tikzpicture}
\end{quote}

%% file: figures/bb-tree-merge-gray.tikz
\begin{tikzpicture}
	\begin{pgfonlayer}{nodelayer}
		\node [style=none] (0) at (0, -0.25) {$=$};
		\node [style=none] (1) at (-3, 0.75) {};
		\node [style=none] (2) at (-4, 0.75) {};
		\node [style=none] (3) at (-3, 1.75) {};
		\node [style=bbox corner] (4) at (-4, 1.75) {};
		\node [style=none] (5) at (-2.25, -2) {};
		\node [style=gray dot] (6) at (-2.25, -1) {};
		\node [style=none] (7) at (-3.5, 1.25) {};
		\node [style=none, blue] (8) at (-4.5, 1.75) {$b$};
		\node [style=gray dot] (9) at (-1.5, -0.25) {};
		\node [style=none, blue] (10) at (-2.5, 1.75) {$c$};
		\node [style=none] (11) at (-2, 0.75) {};
		\node [style=bbox corner] (12) at (-2, 1.75) {};
		\node [style=none] (13) at (-1.5, 1.25) {};
		\node [style=none] (14) at (-1, 0.75) {};
		\node [style=none] (15) at (-1, 1.75) {};
		\node [style=none] (16) at (2, 0.75) {};
		\node [style=none, blue] (17) at (2.5, 1.75) {$c$};
		\node [style=gray dot] (18) at (2.5, -0.5) {};
		\node [style=none] (19) at (3.5, 1.25) {};
		\node [style=none] (20) at (2.5, -2) {};
		\node [style=bbox corner] (21) at (3, 1.75) {};
		\node [style=none] (22) at (2, 1.75) {};
		\node [style=none] (23) at (3, 0.75) {};
		\node [style=none] (24) at (1.5, 1.25) {};
		\node [style=none] (25) at (1, 0.75) {};
		\node [style=none] (26) at (4, 1.75) {};
		\node [style=none] (27) at (4, 0.75) {};
		\node [style=bbox corner] (28) at (1, 1.75) {};
		\node [style=none, blue] (29) at (0.5, 1.75) {$b$};
	\end{pgfonlayer}
	\begin{pgfonlayer}{edgelayer}
		\draw [style=diredge] (6) to (7);
		\draw [style=diredge] (5) to (6);
		\draw [style=bbox edge] (4) to (3.center);
		\draw [style=bbox edge] (3.center) to (1.center);
		\draw [style=bbox edge] (1.center) to (2.center);
		\draw [style=bbox edge] (2.center) to (4);
		\draw [style=diredge] (6) to (9);
		\draw [style=bbox edge] (12) to (15.center);
		\draw [style=bbox edge] (15.center) to (14.center);
		\draw [style=bbox edge] (14.center) to (11.center);
		\draw [style=bbox edge] (11.center) to (12);
		\draw [style=diredge] (9) to (13);
		\draw [style=diredge] (18.center) to (24);
		\draw [style=diredge] (20) to (18.center);
		\draw [style=bbox edge] (28) to (22.center);
		\draw [style=bbox edge] (22.center) to (16.center);
		\draw [style=bbox edge] (16.center) to (25.center);
		\draw [style=bbox edge] (25.center) to (28);
		\draw [style=bbox edge] (21) to (26.center);
		\draw [style=bbox edge] (26.center) to (27.center);
		\draw [style=bbox edge] (27.center) to (23.center);
		\draw [style=bbox edge] (23.center) to (21);
		\draw [style=diredge] (18.center) to (19);
	\end{pgfonlayer}
\end{tikzpicture}

%% file: figures/id-to-wire-gray.tikz
\begin{tikzpicture}
	\begin{pgfonlayer}{nodelayer}
		\node [style=none] (0) at (0, 0) {$=$};
		\node [style=none] (1) at (-1.25, 1) {};
		\node [style=gray dot] (2) at (-1.25, 0) {};
		\node [style=none] (3) at (1, 1) {};
		\node [style=none] (4) at (1, -1) {};
		\node [style=none] (5) at (-1.25, -1) {};
	\end{pgfonlayer}
	\begin{pgfonlayer}{edgelayer}
		\draw [style=diredge] (2) to (1.center);
		\draw [style=diredge] (5.center) to (2);
		\draw [style=diredge] (4.center) to (3.center);
	\end{pgfonlayer}
\end{tikzpicture}

%% file: figures/gen-bialg.tikz
\begin{tikzpicture}
	\begin{pgfonlayer}{nodelayer}
		\node [style=none] (0) at (-2.25, -1.75) {};
		\node [style=none] (1) at (-1.5, -1.5) {...};
		\node [style=none] (2) at (-0.75, -1.75) {};
		\node [style=white dot] (3) at (-1.5, -0.75) {};
		\node [style=none] (4) at (-1.5, 1.5) {...};
		\node [style=none] (5) at (-2.25, 1.75) {};
		\node [style=none] (6) at (-0.75, 1.75) {};
		\node [style=gray dot] (7) at (-1.5, 0.75) {};
		\node [style=none] (8) at (0, 0) {$=$};
		\node [style=gray dot] (9) at (1.75, -1) {};
		\node [style=none] (10) at (2.75, -0.25) {};
		\node [style=gray dot] (11) at (3.25, -1) {};
		\node [style=none] (12) at (2.25, -0.25) {};
		\node [style=none] (13) at (3.25, -1.75) {};
		\node [style=none] (14) at (1.75, -1.75) {};
		\node [style=none] (15) at (2.75, 0.25) {};
		\node [style=white dot] (16) at (1.75, 1) {};
		\node [style=none] (17) at (3.25, 1.75) {};
		\node [style=none] (18) at (1.75, 1.75) {};
		\node [style=none] (19) at (2.25, 0.25) {};
		\node [style=white dot] (20) at (3.25, 1) {};
		\node [style=none] (21) at (2.5, -1.5) {...};
		\node [style=none] (22) at (2.5, 1.5) {...};
		\node [style=none] (23) at (3.25, 0) {...};
		\node [style=none] (24) at (1.75, 0) {...};
	\end{pgfonlayer}
	\begin{pgfonlayer}{edgelayer}
		\draw [style=diredge] (0.center) to (3);
		\draw [style=diredge] (2.center) to (3);
		\draw [style=diredge] (7) to (5.center);
		\draw [style=diredge] (7) to (6.center);
		\draw [style=diredge] (3) to (7);
		\draw [style=diredge] (9) to (10.center);
		\draw [style=diredge] (11) to (12.center);
		\draw [style=diredge] (13.center) to (11);
		\draw [style=diredge] (14.center) to (9);
		\draw [style=diredge] (15.center) to (16);
		\draw [style=diredge] (16) to (18.center);
		\draw [style=diredge] (19.center) to (20);
		\draw [style=diredge] (20) to (17.center);
		\draw [style=diredge, bend left=45, looseness=1.25] (9) to (16);
		\draw [style=diredge, bend right=45, looseness=1.25] (11) to (20);
	\end{pgfonlayer}
\end{tikzpicture}

%% file: figures/gen-bialg-bbox.tikz
\begin{tikzpicture}
	\begin{pgfonlayer}{nodelayer}
		\node [style=none] (0) at (-1.5, -1.75) {};
		\node [style=white dot] (1) at (-1.5, -0.5) {};
		\node [style=none] (2) at (-1.5, 1.75) {};
		\node [style=gray dot] (3) at (-1.5, 0.5) {};
		\node [style=none] (4) at (0, 0) {$=$};
		\node [style=gray dot] (5) at (2.25, -1) {};
		\node [style=none] (6) at (2.25, -2) {};
		\node [style=white dot] (7) at (2.25, 1) {};
		\node [style=none] (8) at (2.25, 2) {};
		\node [style=none] (9) at (-1, 2.25) {};
		\node [style=bbox corner] (10) at (-2, 2.25) {};
		\node [style=none] (11) at (-1, 1.25) {};
		\node [style=none] (12) at (-2, 1.25) {};
		\node [style=none] (13) at (-1, -1.25) {};
		\node [style=bbox corner] (14) at (-2, -1.25) {};
		\node [style=none] (15) at (-1, -2.25) {};
		\node [style=none] (16) at (-2, -2.25) {};
		\node [style=none] (17) at (3, -0.25) {};
		\node [style=bbox corner] (18) at (1.5, -0.25) {};
		\node [style=none] (19) at (3, -2.25) {};
		\node [style=none] (20) at (1.5, -2.25) {};
		\node [style=none] (21) at (3, 2.25) {};
		\node [style=bbox corner] (22) at (1.5, 2.25) {};
		\node [style=none] (23) at (3, 0.25) {};
		\node [style=none] (24) at (1.5, 0.25) {};
		\node [style=none, blue] (25) at (1, 2.25) {$b$};
		\node [style=none, blue] (26) at (1, -0.25) {$c$};
		\node [style=none, blue] (27) at (-2.5, -1.25) {$c$};
		\node [style=none, blue] (28) at (-2.5, 2.25) {$b$};
	\end{pgfonlayer}
	\begin{pgfonlayer}{edgelayer}
		\draw [style=diredge] (0.center) to (1);
		\draw [style=diredge] (3) to (2.center);
		\draw [style=diredge] (1) to (3);
		\draw [style=diredge] (6.center) to (5);
		\draw [style=diredge] (7) to (8.center);
		\draw [style=bbox edge] (10) to (9.center);
		\draw [style=bbox edge] (9.center) to (11.center);
		\draw [style=bbox edge] (11.center) to (12.center);
		\draw [style=bbox edge] (12.center) to (10);
		\draw [style=bbox edge] (14) to (13.center);
		\draw [style=bbox edge] (13.center) to (15.center);
		\draw [style=bbox edge] (15.center) to (16.center);
		\draw [style=bbox edge] (16.center) to (14);
		\draw [style=diredge] (5) to (7);
		\draw [style=bbox edge] (18) to (17.center);
		\draw [style=bbox edge] (17.center) to (19.center);
		\draw [style=bbox edge] (19.center) to (20.center);
		\draw [style=bbox edge] (20.center) to (18);
		\draw [style=bbox edge] (22) to (21.center);
		\draw [style=bbox edge] (21.center) to (23.center);
		\draw [style=bbox edge] (23.center) to (24.center);
		\draw [style=bbox edge] (24.center) to (22);
	\end{pgfonlayer}
\end{tikzpicture}

%% file: figures/arch.tikz
\begin{tikzpicture}[font={\footnotesize}, yscale=1.1, xscale=1.5]
	\begin{pgfonlayer}{nodelayer}
		\node [style=none] (0) at (-10, 3.5) {};
		\node [style=none] (1) at (10, 3.5) {};
		\node [style=none] (2) at (10, 2) {};
		\node [style=none] (3) at (-10, 2) {};
		\node [style=none] (4) at (0, 2.75) {QuantoDerive (Scala)};
		\node [style=none] (5) at (10, -3.25) {};
		\node [style=none] (6) at (3, -1.25) {QuantoCore (ML)};
		\node [style=none] (7) at (-4, 0.5) {};
		\node [style=none] (8) at (-4, -3.25) {};
		\node [style=none] (9) at (10, 0.5) {};
		\node [style=none] (10) at (10, -4.75) {};
		\node [style=none] (11) at (0, -4) {Poly/ML};
		\node [style=none] (12) at (-10, -3.25) {};
		\node [style=none] (13) at (-10, -4.75) {};
		\node [style=none] (14) at (10, -3.25) {};
		\node [style=none] (15) at (-8.75, 2) {};
		\node [style=none] (16) at (-8.75, -3.25) {};
		\node [style=none] (17) at (-10.5, -1.5) {Simprocs};
		\node [style=none] (18) at (-10.5, 0.25) {{\color{red}IDE}};
		\node [style=none] (19) at (-2.25, 0) {Dispatcher};
		\node [style=none] (20) at (-0.5, 0.5) {};
		\node [style=none] (21) at (-0.5, -0.5) {};
		\node [style=none] (22) at (-4, -0.5) {};
		\node [style=none] (23) at (3, -0.5) {};
		\node [style=none] (24) at (3, 0.5) {};
		\node [style=none] (25) at (-0.5, 0.5) {};
		\node [style=none] (26) at (1.25, 0) {Worker};
		\node [style=none] (27) at (-0.5, -0.5) {};
		\node [style=none] (28) at (6.5, -0.5) {};
		\node [style=none] (29) at (3, 0.5) {};
		\node [style=none] (30) at (3, -0.5) {};
		\node [style=none] (31) at (4.75, 0) {Worker};
		\node [style=none] (32) at (3, -0.5) {};
		\node [style=none] (33) at (3, 0.5) {};
		\node [style=none] (34) at (6.5, 0.5) {};
		\node [style=none] (35) at (6.5, -0.5) {};
		\node [style=none] (36) at (10, 0.5) {};
		\node [style=none] (37) at (6.5, 0.5) {};
		\node [style=none] (38) at (10, -0.5) {};
		\node [style=none] (39) at (6.5, 0.5) {};
		\node [style=none] (40) at (6.5, -0.5) {};
		\node [style=none] (41) at (6.5, 0.5) {};
		\node [style=none] (42) at (8.25, 0) {Worker};
		\node [style=none] (43) at (-2.25, 2) {};
		\node [style=none] (44) at (-2.25, 0.5) {};
		\node [style=none] (45) at (1.25, 0.5) {};
		\node [style=none] (46) at (1.25, 2) {};
		\node [style=none] (47) at (4.75, 0.5) {};
		\node [style=none] (48) at (4.75, 2) {};
		\node [style=none] (49) at (8.25, 0.5) {};
		\node [style=none] (50) at (8.25, 2) {};
		\node [style=none] (51) at (-4.75, 1.25) {{\color{blue}JSON Protocol}};
		\node [style=none] (52) at (10, 3.5) {};
		\node [style=none] (53) at (-10, 3.5) {};
		\node [style=none] (54) at (-0.75, -3) {};
		\node [style=none] (55) at (-3.75, -2) {};
		\node [style=none] (56) at (-3.75, -2) {};
		\node [style=none] (57) at (-0.75, -2) {};
		\node [style=none] (58) at (-3.75, -3) {};
		\node [style=none] (59) at (-3.75, -3) {};
		\node [style=none] (60) at (8.25, -2.5) {Theories};
		\node [style=none] (61) at (2.75, -2) {};
		\node [style=none] (62) at (-0.25, -3) {};
		\node [style=none] (63) at (1.25, -2.5) {Matching};
		\node [style=none] (64) at (-0.25, -2) {};
		\node [style=none] (65) at (2.75, -3) {};
		\node [style=none] (66) at (-0.25, -2) {};
		\node [style=none] (67) at (6.75, -2) {};
		\node [style=none] (68) at (6.25, -3) {};
		\node [style=none] (69) at (6.75, -2) {};
		\node [style=none] (70) at (9.75, -2) {};
		\node [style=none] (71) at (9.75, -3) {};
		\node [style=none] (72) at (6.75, -3) {};
		\node [style=none] (73) at (6.75, -3) {};
		\node [style=none] (74) at (3.25, -2) {};
		\node [style=none] (75) at (4.75, -2.5) {Rewriting};
		\node [style=none] (76) at (3.25, -2) {};
		\node [style=none] (77) at (6.25, -2) {};
		\node [style=none] (78) at (6.25, -3) {};
		\node [style=none] (79) at (3.25, -3) {};
		\node [style=none] (80) at (3.25, -3) {};
		\node [style=none] (81) at (-10.5, -0.5) {{\color{red}Protocol}};
		\node [style=none] (82) at (-4, 0.5) {};
		\node [style=none] (83) at (-2.25, -2.5) {Graphs};
		\node [style=none] (84) at (-7.75, 2) {};
		\node [style=none] (85) at (-7.75, -3.25) {};
	\end{pgfonlayer}
	\begin{pgfonlayer}{edgelayer}
		\draw [style=simple] (3.center) to (0.center);
		\draw [style=simple] (0.center) to (1.center);
		\draw [style=simple] (1.center) to (2.center);
		\draw [style=simple] (2.center) to (3.center);
		\draw [style=simple] (8.center) to (7.center);
		\draw [style=simple] (7.center) to (9.center);
		\draw [style=simple] (9.center) to (5.center);
		\draw [style=simple] (13.center) to (12.center);
		\draw [style=simple] (12.center) to (14.center);
		\draw [style=simple] (14.center) to (10.center);
		\draw [style=simple] (10.center) to (13.center);
		\draw [style=directed, draw=red] (15.center) to (16.center);
		\draw [style=simple] (20.center) to (21.center);
		\draw [style=simple] (22.center) to (21.center);
		\draw [style=simple] (27.center) to (23.center);
		\draw [style=simple] (33.center) to (30.center);
		\draw [style=simple] (32.center) to (28.center);
		\draw [style=simple] (37.center) to (40.center);
		\draw [style=simple] (35.center) to (38.center);
		\draw [style=directed, draw=blue] (43.center) to node{} (44.center);
		\draw [style=directed, draw=blue] (45.center) to node{} (46.center);
		\draw [style=directed, draw=blue] (47.center) to node{} (48.center);
		\draw [style=directed, draw=blue] (49.center) to node{} (50.center);
		\draw [style=simple] (58.center) to (56.center);
		\draw [style=simple] (54.center) to (57.center);
		\draw [style=simple] (55.center) to (57.center);
		\draw [style=simple] (58.center) to (54.center);
		\draw [style=simple] (62.center) to (66.center);
		\draw [style=simple] (65.center) to (61.center);
		\draw [style=simple] (64.center) to (61.center);
		\draw [style=simple] (62.center) to (65.center);
		\draw [style=simple] (72.center) to (67.center);
		\draw [style=simple] (71.center) to (70.center);
		\draw [style=simple] (69.center) to (70.center);
		\draw [style=simple] (73.center) to (71.center);
		\draw [style=simple] (79.center) to (74.center);
		\draw [style=simple] (78.center) to (77.center);
		\draw [style=simple] (76.center) to (77.center);
		\draw [style=simple] (80.center) to (78.center);
		\draw [style=directed, draw=red] (85.center) to (84.center);
	\end{pgfonlayer}
\end{tikzpicture}

%% file: quanto-cade.bbl
\begin{thebibliography}{10}

\bibitem{AC2004}
Samson Abramsky and Bob Coecke.
\newblock A categorical semantics of quantum protocols.
\newblock In {\em LICS 2004}, pages 415--425. IEEE Computer Society, 2004.

\bibitem{Baez2014a}
John~C. Baez and Jason Erbele.
\newblock Categories in control, 2014.
\newblock arXiv:1405.6881.

\bibitem{pawel-bialg}
F.~Bonchi, P.~Soboci\'{n}ski, and F.~Zanasi.
\newblock Interacting bialgebras are frobenius.
\newblock In {\em FoSSaCS `14}, 2014.

\bibitem{Bonchi2015}
Filippo Bonchi, Pawe{\l} Soboci\'{n}ski, and Fabio Zanasi.
\newblock Full abstraction for signal flow graphs.
\newblock In {\em Principles of Programming Languages, POPL`15.}, 2015.

\bibitem{Bob-Coecke:2010fk}
Stephen Clark, Bob Coecke, and Mehrnoosh Sadrzadeh.
\newblock Mathematical foundations for a compositional distributed model of
  meaning.
\newblock {\em Linguistic Analysis}, 36(1-4), 2011.

\bibitem{CD2}
Bob Coecke and Ross Duncan.
\newblock Interacting quantum observables: categorical algebra and
  diagrammatics.
\newblock {\em New Journal of Physics}, 13(4):043016, 2011.

\bibitem{DanosEtAl-LICS2010}
Vincent Danos, J\'er\^ome Feret, Walter Fontana, Russell Harmer, and Jean
  Krivine.
\newblock Abstracting the differential semantics of rule-based models: exact
  and automated model reduction.
\newblock In {\em LICS}, 2010.

\bibitem{DK}
Lucas Dixon and Aleks Kissinger.
\newblock Open-graphs and monoidal theories.
\newblock {\em Mathematical Structures in Computer Science}, 23:308--359, 2013.

\bibitem{Duncan:2013lr}
Ross Duncan and Maxime Lucas.
\newblock Verifying the steane code with quantomatic.
\newblock In {\em Quantum Physics and Logic 2013}, 2013.

\bibitem{mbqc_circuit}
Ross Duncan and Simon Perdrix.
\newblock Rewriting measurement-based quantum computations with generalised
  flow.
\newblock In {\em Automata, Languages and Programming}. 2010.

\bibitem{mehrnoosh_experiment}
Edward Grefenstette and Mehrnoosh Sadrzadeh.
\newblock Experimental support for a categorical compositional distributional
  model of meaning.
\newblock In {\em Proceedings of the Conference on Empirical Methods in Natural
  Language Processing}, 2011.

\bibitem{tinker14}
Gudmund Grov, Aleks Kissinger, and Yuhui Lin.
\newblock Tinker, tailor, solver, proof.
\newblock In {\em UITP}, 2014.

\bibitem{Hillebrand:2011qy}
Anne Hillebrand.
\newblock Quantum protocols involving multiparticle entanglement and their
  representations in the {ZX}-calculus.
\newblock Master's thesis, Oxford University, 2011.

\bibitem{IsaCosy}
Moa Johansson, Lucas Dixon, and Alan Bundy.
\newblock Conjecture synthesis for inductive theories.
\newblock {\em Journal of Automated Reasoning}, 47(3):251--289, 2011.

\bibitem{JS}
André Joyal and Ross Street.
\newblock The geometry of tensor calculus, {I}.
\newblock {\em Advances in Mathematics}, 88(1):55 -- 112, 1991.

\bibitem{QuantoCosy}
A.~Kissinger.
\newblock Synthesising graphical theories.
\newblock 2012.
\newblock {arXiv:1202.6079}.

\bibitem{Kissinger:2012lr}
Aleks Kissinger.
\newblock {\em Pictures of Processes: Automated Graph Rewriting for Monoidal
  Categories and Applications to Quantum Computing}.
\newblock PhD thesis, Oxford, 2012.

\bibitem{pattern_graphs}
Aleks Kissinger, Alex Merry, and Matvey Soloviev.
\newblock Pattern graph rewrite systems.
\newblock In {\em Proceedings of DCM}, 2012.

\bibitem{Noncomm-bb}
Aleks Kissinger and David Quick.
\newblock Tensors, !-graphs, and non-commutative quantum structures.
\newblock In {\em QPL 2014}, volume 172 of {\em EPTCS}, pages 56--67, 2014.

\bibitem{trivial_overlap}
Aleks Kissinger and Vladimir Zamdzhiev.
\newblock !-graphs with trivial overlap are context-free.
\newblock In {\em Proceedings Graphs as Models, GaM 2015, London, UK, 11-12
  April 2015.}, 2015.

\bibitem{Mellies:2014aa}
Paul-Andr{\'e} Melli{\`e}s.
\newblock Local states in string diagrams.
\newblock In {\em RTA-TLCA}, 2014.

\bibitem{Merry:2013aa}
Alexander Merry.
\newblock {\em Reasoning with !-graphs}.
\newblock PhD thesis, Oxford University, 2013.

\bibitem{gudmund:11c}
Greg Michaelson and Gudmund Grov.
\newblock Reasoning about multi-process systems with the box calculus.
\newblock In {\em Lectures from Central European Functional Programming
  School}. Springer, 2011.

\bibitem{Penrose}
R.~Penrose.
\newblock Applications of negative dimensional tensors.
\newblock In {\em Combinatorial Mathematics and its Applications}, pages
  221--244. Academic Press, 1971.

\bibitem{groove}
Arend Rensink.
\newblock {The GROOVE Simulator: A Tool for State Space Generation}.
\newblock In {\em Applications of Graph Transformations with Industrial
  Relevance}. 2004.

\bibitem{progress}
Andy Sch{\"u}rr.
\newblock {PROGRESS: A VHL-language based on graph grammars}.
\newblock In {\em Graph Grammars and Their Application to Computer Science}.
  1991.

\bibitem{Sobocinski:2010aa}
P.~Sobocinski.
\newblock Representations of petri net interactions.
\newblock In {\em CONCUR 2010 - Concurrency Theory}, volume 6269 of {\em LNCS},
  pages pp 554--568. Springer, 2010.

\bibitem{agg}
Gabriele Taentzer.
\newblock {AGG: A Graph Transformation Environment for Modeling and Validation
  of Software}.
\newblock In {\em Applications of Graph Transformations with Industrial
  Relevance}. 2004.

\end{thebibliography}
